\documentclass[preprint,aps,amsmath,showpacs,showkeys,nofootinbib, 
superscriptaddress]{revtex4} 
\usepackage{graphicx} 
 
\begin{document} 

\title{\textbf{$K^-$- nuclear states: Binding energies and widths}} 

\author{J.~Hrt\'{a}nkov\'{a}} 
\email{hrtankova@ujf.cas.cz} 
\affiliation{Nuclear Physics Institute, 25068 \v{R}e\v{z}, Czech Republic} 

\author{J.~Mare\v{s}} 
\email{mares@ujf.cas.cz} 
\affiliation{Nuclear Physics Institute, 25068 \v{R}e\v{z}, Czech Republic} 

\date{\today} 

\begin{abstract} 
$K^-$ optical potentials relevant to calculations of $K^-$ nuclear quasi-bound states were developed  
within several chiral meson-baryon coupled-channel interaction models. The applied models yield quite different $K^-$ binding energies and widths. Then, the $K^-$ multinucleon interactions were incorporated by a phenomenological optical potential 
fitted recently to kaonic atom data.     
Though the applied $K^-$ interaction models differ significantly in the $K^-N$ subthreshold region, 
our self-consistent calculations of kaonic nuclei across the periodic table lead to conclusions 
valid quite generally. Due to $K^-$ multinucleon absorption in the nuclear medium 
the calculated widths of $K^-$ nuclear states are sizable, $\Gamma_{K^-} \geq 90$~MeV, and exceed 
substantially their binding energies in all considered nuclei.  
\end{abstract} 



\maketitle 

\section{Introduction}
\label{intro}

The near-threshold ${\bar K}N$ attraction seems to be strong enough to bind the antikaon in the nuclear medium 
and form a kaonic nucleus~\cite{ayPRC, yaPLB, W10, Shev_review}.  
However, strong absorption of $K^-$ in nuclear matter, as well as in-medium modifications and distinct energy 
dependence of the $K^-N$ scattering amplitudes attributed to the $\Lambda(1405)$ resonance could call this 
presumption into question and thus have to be carefully accounted for in relevant calculations.   

Unique information allowing us to fix the $K^- p$ interaction at and above threshold is provided by low-energy 
${\bar K}N$ scattering data (summarized e.g. in Ref.~\cite{kmnlo}), threshold branching ratios~\cite{mNPB}, and in particular, strong interaction energy shift and width of kaonic hydrogen atom~\cite{sidhharta}. 
The $K^- n$ interaction is much poorly determined due to the lack of sufficiently accurate data. Considerably 
less is known about the $K^- N$ interaction below threshold. Information about the subthreshold interaction of 
$K^-$ with nucleons comes from the analyses of $\pi\Sigma$ spectra in the region of $\Lambda(1405)$ and especially 
from the measurement of energy shifts and widths of $K^-$ atomic states throughout the periodic table~\cite{bfg97,fg07}. 

\begin{figure}[t!]
\begin{center}
\includegraphics[width=0.45\textwidth]{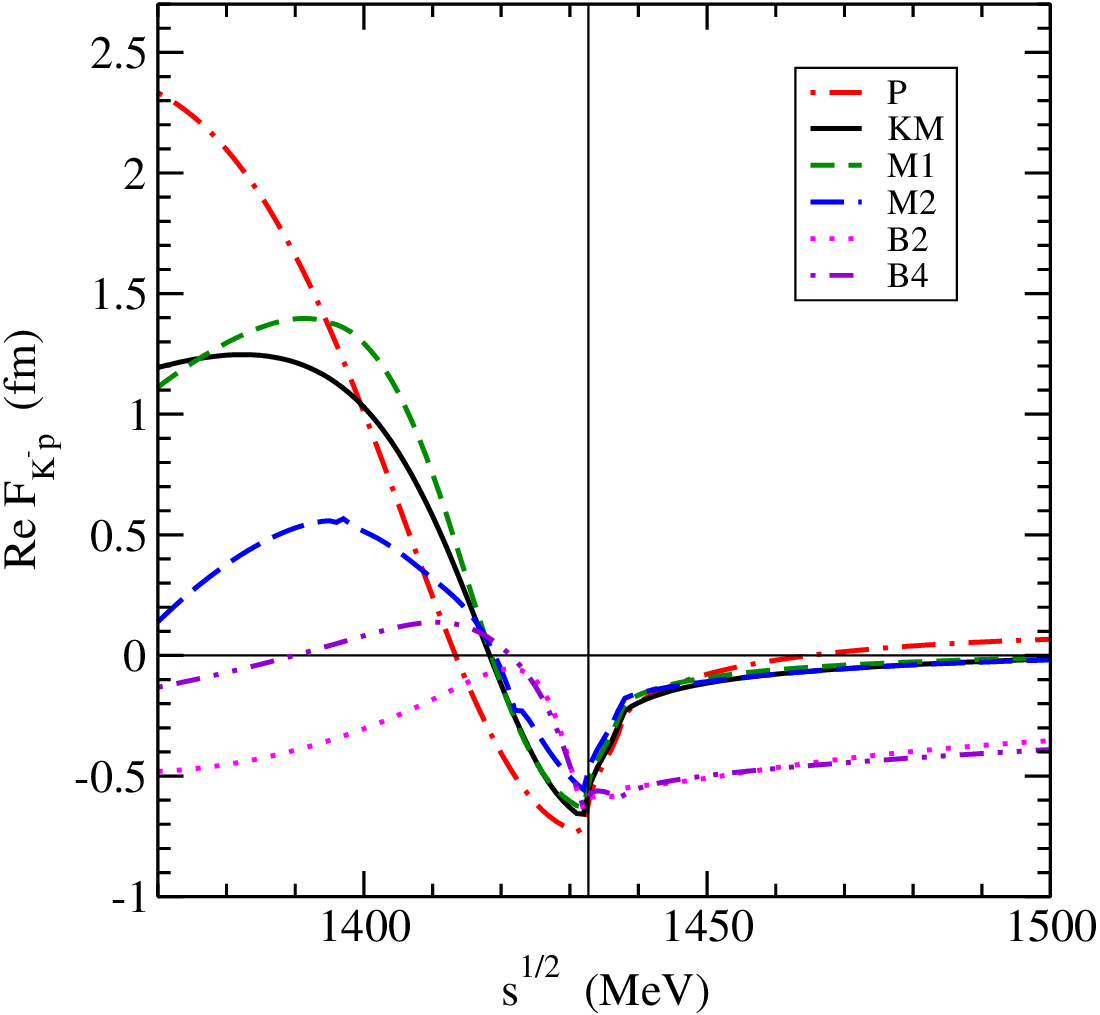} \hspace{10pt}
\includegraphics[width=0.45\textwidth]{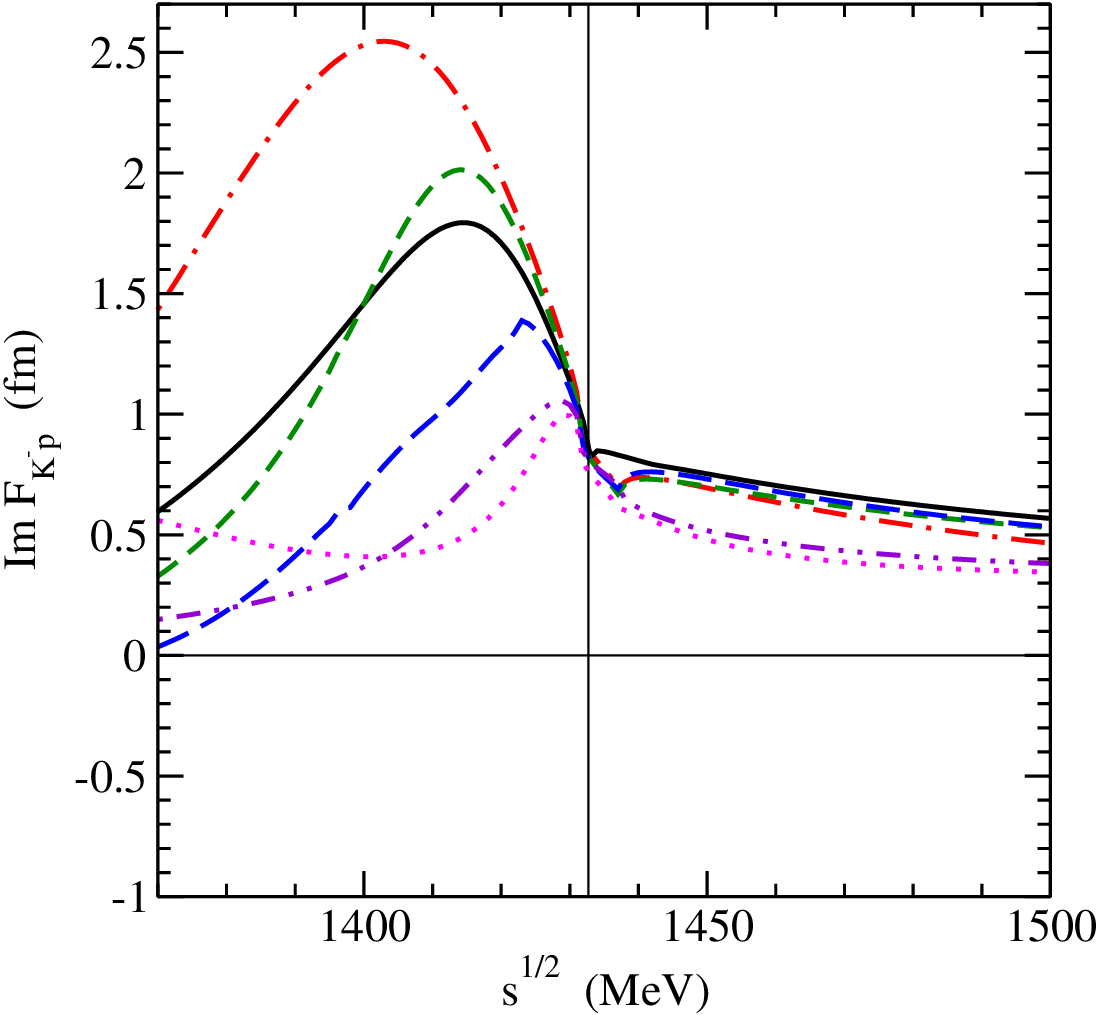} \\[10pt]
\includegraphics[width=0.45\textwidth]{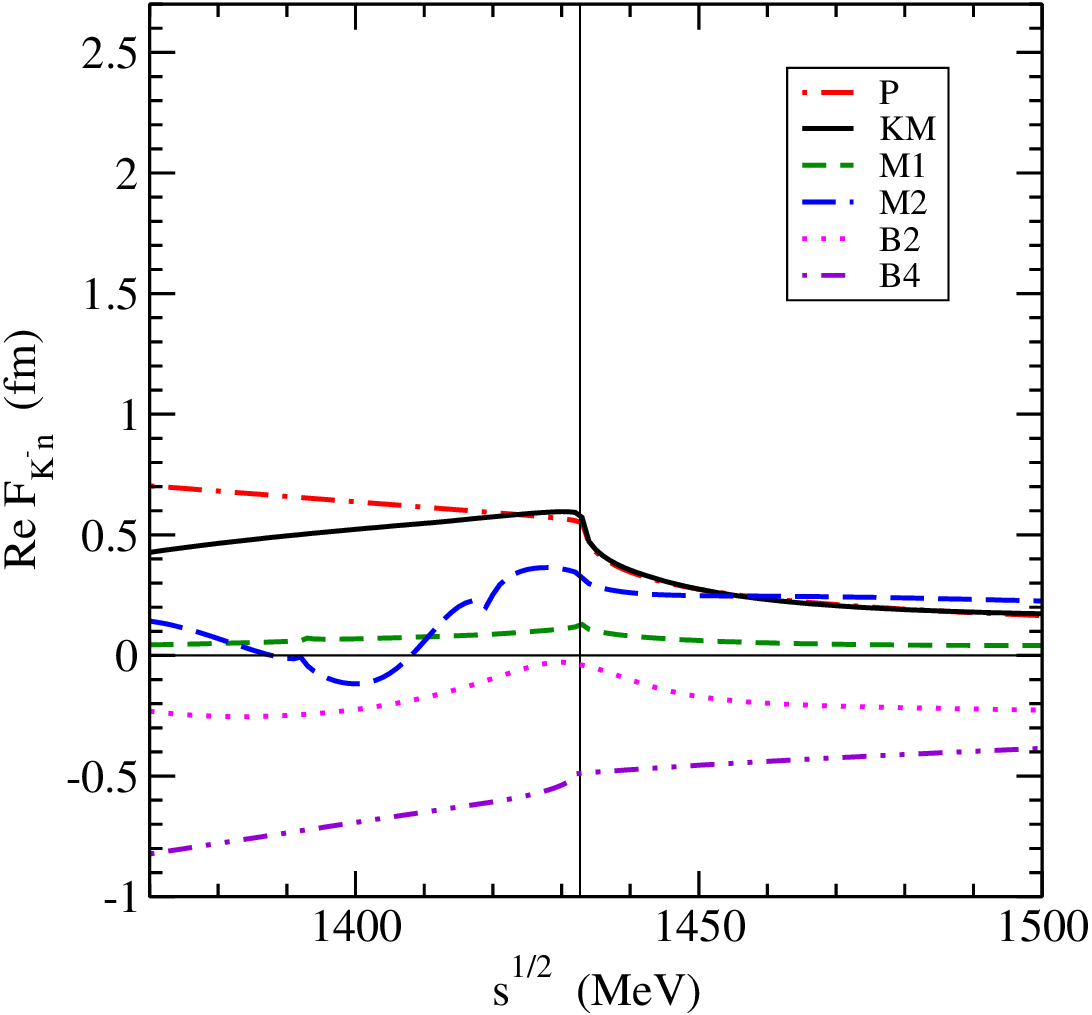} \hspace{10pt}
\includegraphics[width=0.45\textwidth]{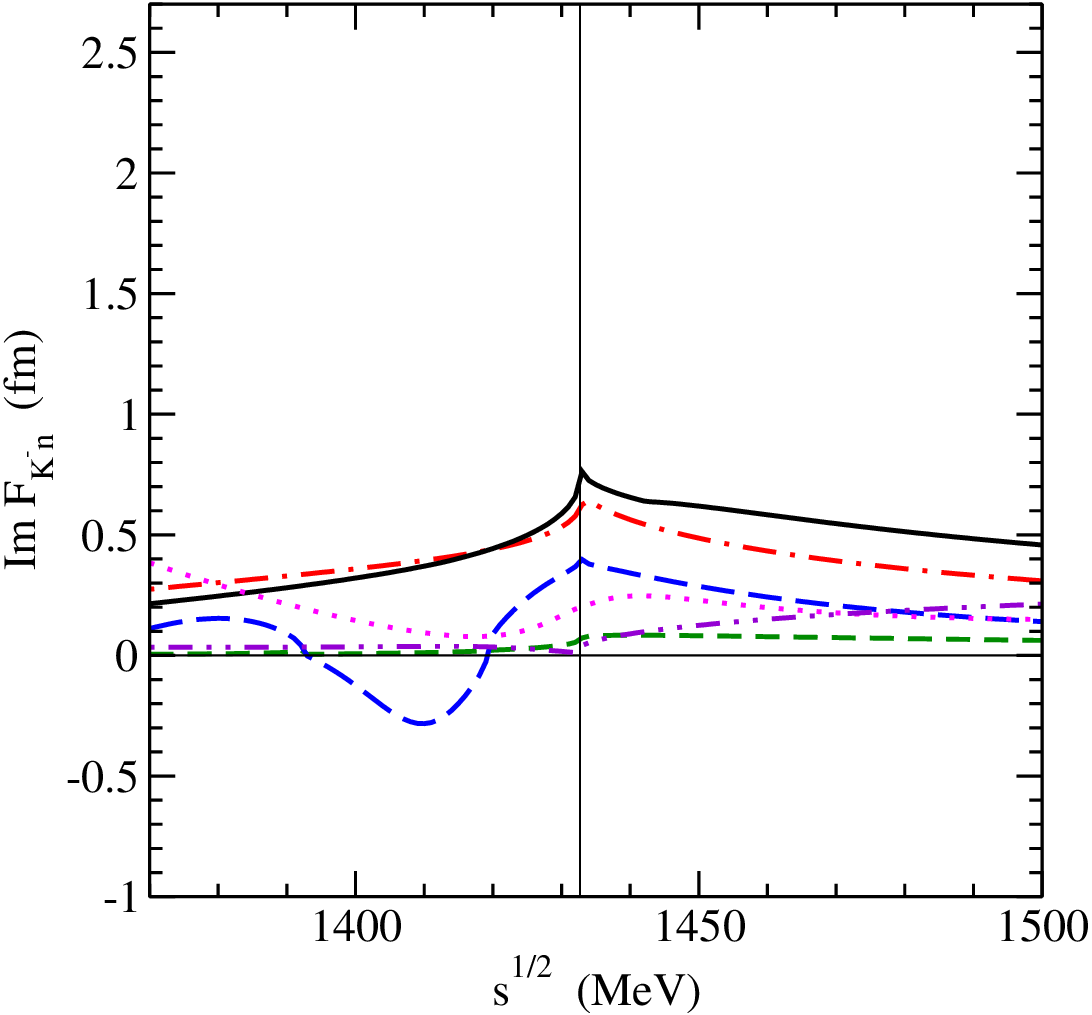}
\end{center}
\caption{Energy dependence of real (left) and imaginary (right) parts of free-space $K^-p$ (top) and $K^-n$ (bottom) amplitudes in considered chiral models (see text for details). Thin vertical lines mark threshold energies.}
\label{fig.:KpnFree}
\end{figure}

The theoretical description of the $K^-N$ interaction is currently provided by chirally-motivated meson-baryon 
interaction models. Parameters of these models are tuned to reproduce the above low-energy $K^-N$ observables. 
In the present study, the free-space $K^-N$ scattering amplitudes derived within various chiral 
SU(3) meson-baryon coupled-channel interaction models: 
Prague (P) \cite{pnlo}, Kyoto-Munich (KM) \cite{kmnlo}, Murcia (M1 and M2) \cite{m}, and Bonn (B2 and B4) \cite{b} 
are used to construct the kaon self-energy operator $\Pi_{K^-}$. 
The free $s$-wave scattering amplitudes $F_{K^-p}(\sqrt{s})$ and  $F_{K^-n}(\sqrt{s})$ considered in this 
work are shown in Fig.~\ref{fig.:KpnFree}. Being constrained by the data, the $F_{K^-p}(\sqrt{s})$ amplitudes 
(Fig.~\ref{fig.:KpnFree} top) agree with each other at threshold and, 
except the Bonn model amplitudes, also above threshold. The form of B2 and B4 amplitudes deviates from the others 
because higher partial waves were included in the Bonn model fits. All the $K^- p$ amplitudes differ considerably 
below threshold, which implies the region relevant 
for $K^-$-nuclear bound-state calculations. Moreover, they are significantly energy-dependent below threshold due to existence of $\Lambda(1405)$ resonance which is dynamically generated in these models. It is thus important to evaluate the $K^-$-nucleus potential self-consistently~\cite{cfggmPLB, cfggmPRC11}. The $K^- n$ amplitudes (Fig.~\ref{fig.:KpnFree} bottom) differ appreciably from each other in the entire energy range considered here. Figure~\ref{fig.:KpnFree} illustrates significant model dependence of the input scattering amplitudes. As a result, 
binding energies $B_{K^-}$ and widths $\Gamma_{K^-}$ of kaonic nuclear states calculated within the above $K^-N$ interaction models are expected to differ substantially from each other.  
 
The implications of self-consistent treatment of energy dependence of chirally-inspired $K^-N$ amplitudes 
near threshold for calculations of $K^-$-nuclear states were discussed in Ref.~\cite{gmNPA}.   
Due to a sizable downward energy shift towards $\pi \Sigma$ threshold, the $K^-$ potential constructed within 
the P model yields relatively small $K^-$ widths because only the $K^-$ absorption on a 
single-nucleon, $K^-N\rightarrow \pi Y$ ($Y= \Lambda, \Sigma$), is involved in this 
model~\cite{cfggmPLB, cfggmPRC11, gmNPA}. 
In nuclear medium,  $K^-$ multinucleon interactions, such as $K^-NN \rightarrow YN$  take place as well~\cite{fgb93, fgNPA, fgNPA16} 
and should thus be considered in any realistic study of $K^-$-nuclear quasi-bound 
states. Indeed, recent analyses of kaonic atoms have confirmed that a phenomenological term 
representing $K^-$ multinucleon processes has to be added to the optical potential constructed from 
in-medium chirally motivated $K^-N$ amplitudes in order to achieve good fit to the data~\cite{fgNPA, fgNPA16}.
In Refs. ~\cite{cfggmPLB, cfggmPRC11, gmNPA}, the $K^-NN$ absorption was included 
using a phenomenological potential and as a consequence, the $K^-$ widths increased and became comparable 
with $K^-$ binding energies.      
Although the chiral $K^-N$ interaction models do not involve the $K^-$ multinucleon 
processes explicitly, 
Sekihara \emph{et al.}~\cite{sjPRC12} derived non-mesonic $K^-$ interaction channels within 
a chiral unitary approach for the $s$-wave ${\bar K}N$ amplitude and calculated the ratio of mesonic to 
non-mesonic $K^-$ absorption at rest in nuclear matter. 
The experimental information about this ratio comes from bubble chamber experiments \cite{bubble1, bubble2, bubble3}. 
Recently, Friedman and Gal have supplemented the $K^-$ single-nucleon potential constructed 
from several chiral $K^-N$ amplitude models by a phenomenological term representing the $K^-$ multinucleon 
interactions and fitted its parameters to kaonic atom data for each meson-baryon interaction model 
separately~\cite{fgNPA16}. Moreover, they confronted the total $K^-$ optical potential with experimental 
fractions of $K^-$ absorption at rest. They found that only the P and KM models supplemented 
by the $K^-$ multinucleon potential are able to reproduce both experimental constraints simultaneously. 
These two models were recently used in calculations of $K^-$ quasi-bound states~\cite{hmPLB17} and the $K^-$ 
multinucleon interactions were found to cause radical increase of the widths of $K^-$-nuclear states. 
 
In this work, we apply all six chirally-motivated meson-baryon coupled-channel interaction models considered 
in Ref.~\cite{fgNPA16} to calculations of $K^-$- nuclear quasi bound states, aiming at exploring model 
dependence of predicted $K^-$ binding energies and widths. Then we supplement the $K^-$ single-nucleon potential 
by a corresponding phenomenological optical potential describing the $K^-$ multinucleon interactions in order to 
study in detail their impact on $K^-$ binding energies and widths. 
Unlike previous calculations, we consider various $K^- N$ interaction models presented in recent years. Most of them  
were never applied in such studies before. We perform unique calculations of kaonic nuclear quasi-bound states 
using the $K^-$-nuclear potentials containing both $K^-$ single-nucleon and multinucleon interactions which were 
fitted to available data for each meson-baryon interaction model. 

The paper is organized as follows. In Section~\ref{sec-1} we present construction of the 
in-medium $K^- N$ amplitudes from the free-space amplitudes derived within chirally-inspired coupled-channel 
models of meson-baryon interactions. We introduce a self-consistent scheme for treating energy dependence of 
these amplitudes and derive for each interaction model a relevant $K^-$-nuclear potential. 
We discuss results of our calculations of $K^-$-nuclear quasi-bound states using these potentials. 
In Section~\ref{sec-2}, we present phenomenological potentials describing $K^-$ multinucleon interactions and  
explore their impact on the widths and binding energies of kaonic nuclear quasi-bound states. 
A brief summary is given in Section~\ref{sec-3}.

\section{Chirally-motivated $K^-$ nuclear potentials}
\label{sec-1}

The binding energies $B_{K^-}$ and widths $\Gamma_{K^-}$ of $K^-$-nuclear quasi-bound states are determined by solving 
self-consistently the Klein-Gordon equation 
\begin{equation}\label{KG}
 \left[ \vec{\nabla}^2  + \tilde{\omega}_{K^-}^2 -m_{K^-}^2 -\Pi_{K^-}(\omega_{K^-},\rho) \right]\phi_{K^-} = 0~,
\end{equation}
where $\tilde{\omega}_{K^-} = m_{K^-} - B_{K^-} -{\rm i}\Gamma_{K^-}/2 -V_C= \omega_{K^-} - V_C$,  $m_{K^-}$ is the $K^-$ mass,  $V_C$ is 
the Coulomb potential introduced via the minimal substitution \cite{kkwPRL90}, and  $\rho$ is the nuclear density distribution. The energy- and density-dependent kaon self-energy operator 
$\Pi_{K^-}$ describes $K^-$ interactions with the nuclear medium.     

The self-energy operator $\Pi_{K^-}$ in Eq.~\eqref{KG} is constructed in a ``$t\rho$'' form 
with the in-medium amplitudes derived from the chirally-motivated $K^- N$ scattering amplitudes presented in Fig.~\ref{fig.:KpnFree}. It is expressed as 
\begin{equation}\label{piK}
\Pi_{K^-} = 2\text{Re}( {\omega}_{K^-})V_{K^-}^{(1)}=-4\pi \frac{\sqrt{s}}{m_N}\left(F_0\frac{1}{2}\rho_p + F_1\left(\frac{1}{2}\rho_p+\rho_n\right)\right)~,
\end{equation}
where $F_0$ and $F_1$ are the isospin 0 and 1 $s$-wave in-medium amplitudes, respectively, $\sqrt{s}$ is the total energy of the $K^-N$ system , $m_N$ is the nucleon mass, and $V_{K^-}^{(1)}$ stands for the (single-nucleon) $K^-$-nucleus optical potential. The kinematical factor $\sqrt{s}/{m_N}$ comes from transforming amplitudes from the two-body cm frame to the lab frame.  
The $\rho_p$ and $\rho_n$ denote proton and neutron density distributions, respectively, in a given core nucleus obtained within 
the relativistic mean-field model NL-SH~\cite{nlsh}. We consider static nuclear density distribution, which means that core polarization effects 
are not included in our calculations. The polarization effects are $A$-dependent -- for instance within the P model, they increase 
$B_{K^-}$ by $\approx 6$~MeV in Li, by $\leq 2$~MeV in Ca, and by $\leq 0.5$~MeV in Pb \cite{gmNPA}. In any case, the role of the nuclear polarization 
is less pronounced than the model dependence.     

The modifications of the free-space amplitudes due to Pauli principle in the medium are accounted for by using the multiple scattering approach (WRW)~\cite{wrw}. The in-medium amplitudes $F_0$ and $F_1$ are then given in the following form:
\begin{equation}\label{Eq.:in-med amp}
F_{1}=\frac{F_{K^-n}(\sqrt{s})}{1+\frac{1}{4}\xi_k \frac{\sqrt{s}}{m_N} F_{K^-n}(\sqrt{s}) \rho}~, \quad F_{0}=\frac{[2F_{K^-p}(\sqrt{s})-F_{K^-n}(\sqrt{s})]}{1+\frac{1}{4}\xi_k \frac{\sqrt{s}}{m_N}[2F_{K^-p}(\sqrt{s}) - F_{K^-n}(\sqrt{s})] \rho}~,
\end{equation}
where
\begin{equation}\label{ksi}
 \xi_k=\frac{9\pi}{p_{\rm F}^2}\,4 I_q,\;\;\;\;\; I_q = \int_0^{\infty} \frac{dt}{t} \exp(iqt)j_1^2(t).
\end{equation}
Here, $p_{\rm F}$ is the Fermi momentum corresponding to density $\rho = 2p_{\rm F}^3/(3\pi^2)$, $j_1(t)$ is the spherical Bessel function and $q= \sqrt{\omega_{K^-}^2-m_{K^-}^2}/p_{\rm F}$. The integral $I_q$ in Eq.~\eqref{ksi} can be evaluated analytically as \cite{fgNPA16}
\begin{equation}
 4 I_q = 1- \frac{q^2}{6} + \frac{q^2}{4}\left(2+ \frac{q^2}{6}\right) \ln \left(1+\frac{4}{q^2}\right) - \frac{4}{3}q\left( \frac{\pi}{2} - \arctan(q/2)\right).
\end{equation}

\begin{figure}[t!]
\begin{center}
\includegraphics[width=0.45\textwidth]{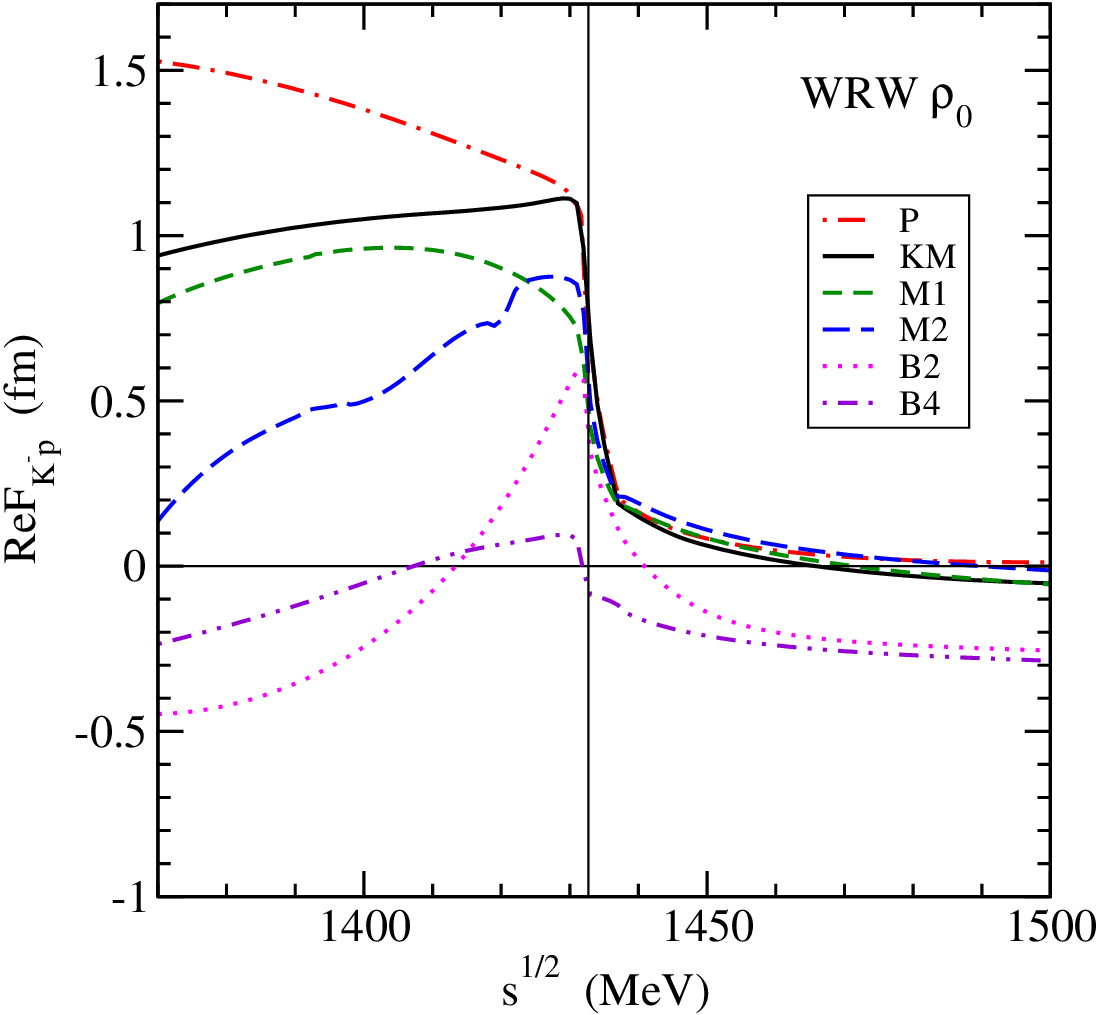} \hspace{10pt}
\includegraphics[width=0.45\textwidth]{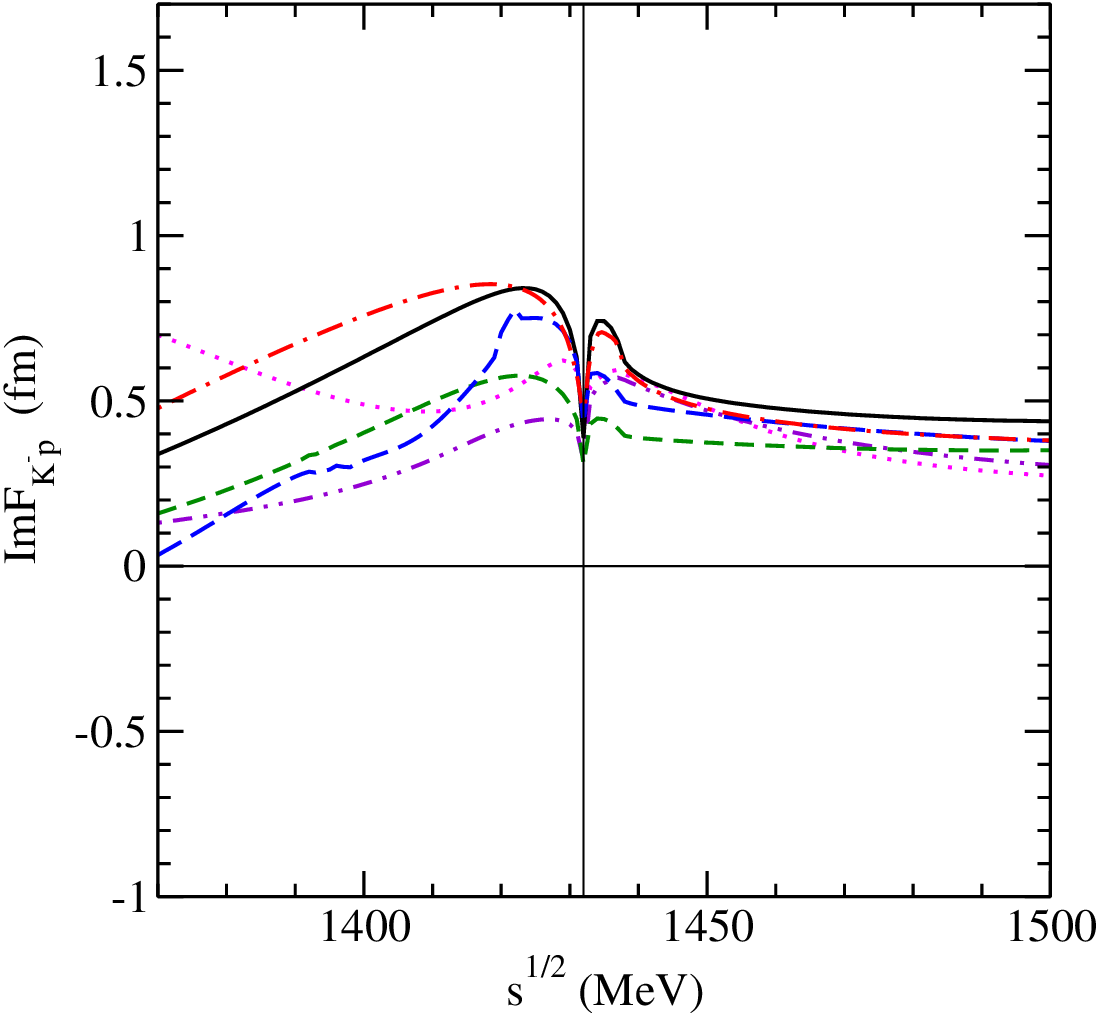} \\[10pt]
\includegraphics[width=0.45\textwidth]{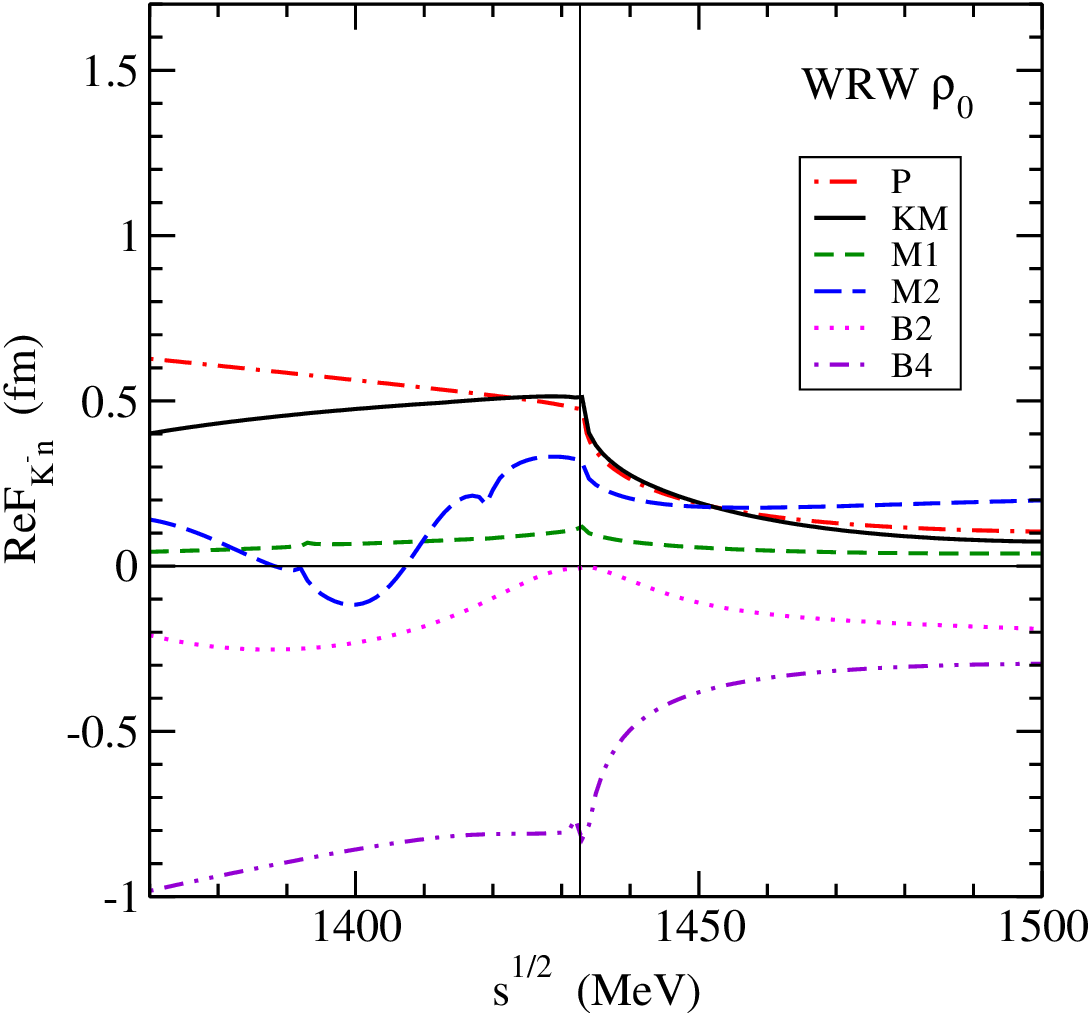} \hspace{10pt}
\includegraphics[width=0.45\textwidth]{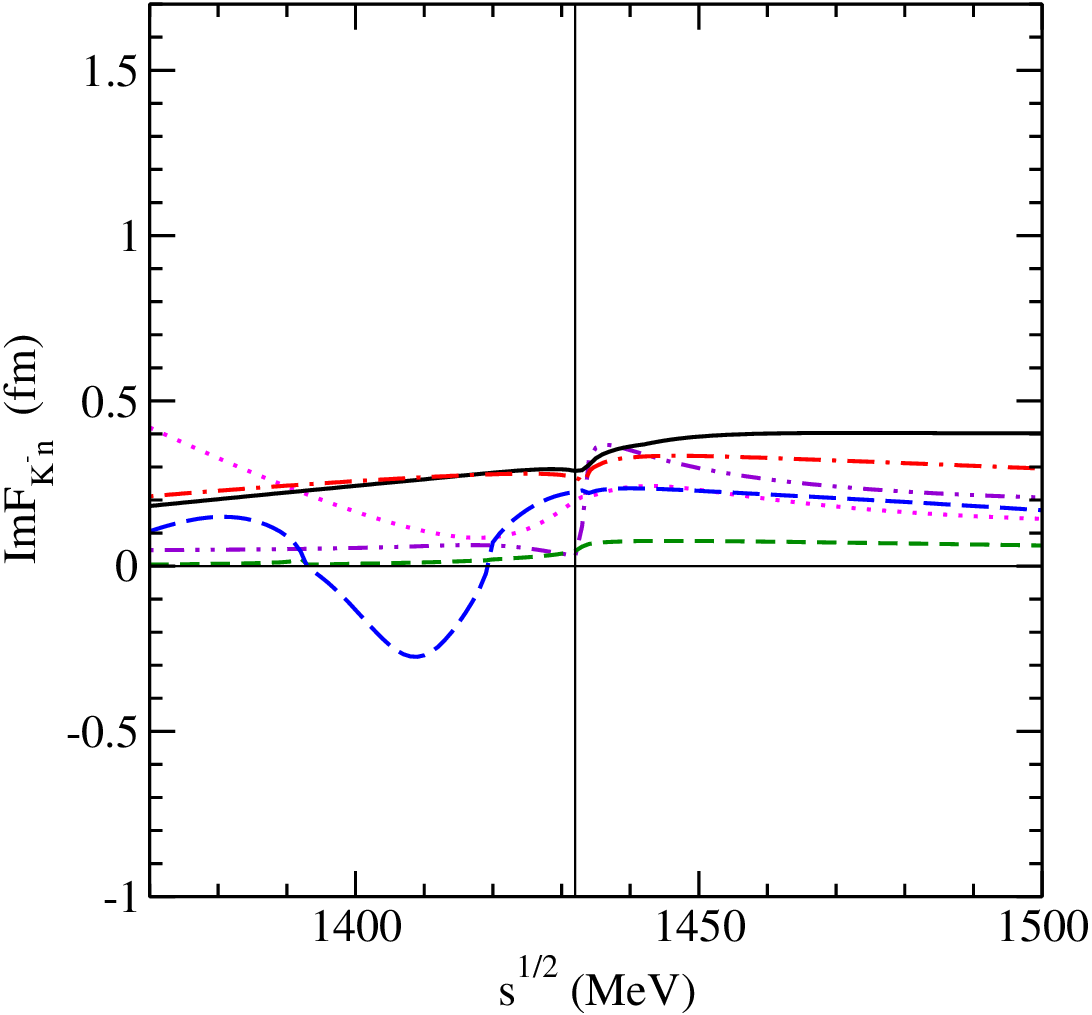}
\end{center}
\caption{Energy dependence of real (left) and imaginary (right) parts of WRW modified $K^-p$ (top) and $K^-n$ (bottom) amplitudes at $\rho_0=0.17$~fm$^{-3}$ in considered models. Thin vertical lines mark threshold energies.}
\label{fig.:inmedamp}
\end{figure}
In Fig.~\ref{fig.:inmedamp}, we present the $K^-p$ and $K^-n$ amplitudes in the considered models, modified by the WRW procedure at 
saturation density $\rho_0=0.17$~fm$^{-3}$ plotted as a function of energy. It follows from comparison with Fig.~\ref{fig.:KpnFree} that the $K^-p$ amplitudes are affected significantly by Pauli correlations: The real part of the amplitudes becomes attractive in the entire energy region below threshold (except the B2 and B4 models) and the imaginary part is considerably lowered below threshold. On the other hand, the $K^-n$ amplitudes are modified by Pauli correlations only 
moderately. 

\begin{figure}[t]
\begin{center}
\includegraphics[width=0.45\textwidth]{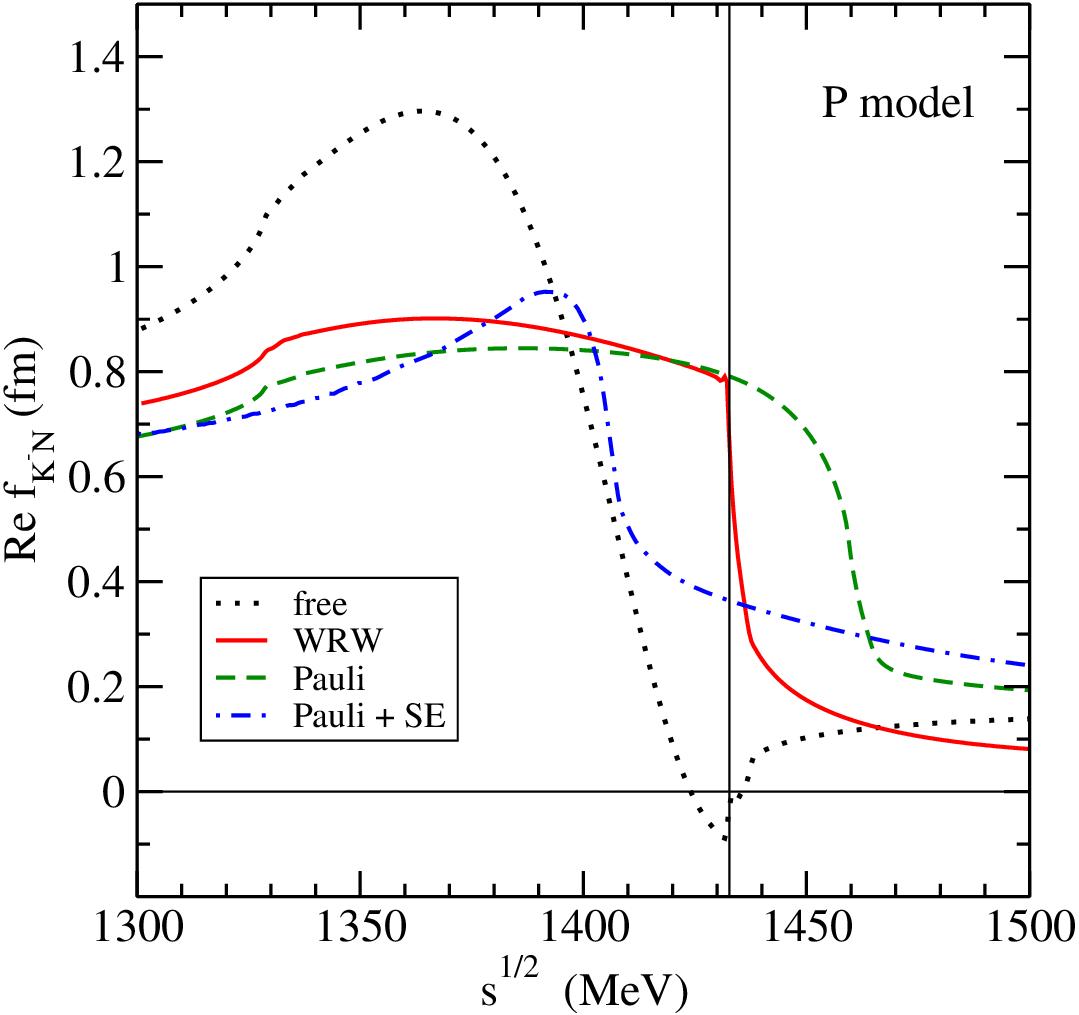} \hspace{20pt}
\includegraphics[width=0.45\textwidth]{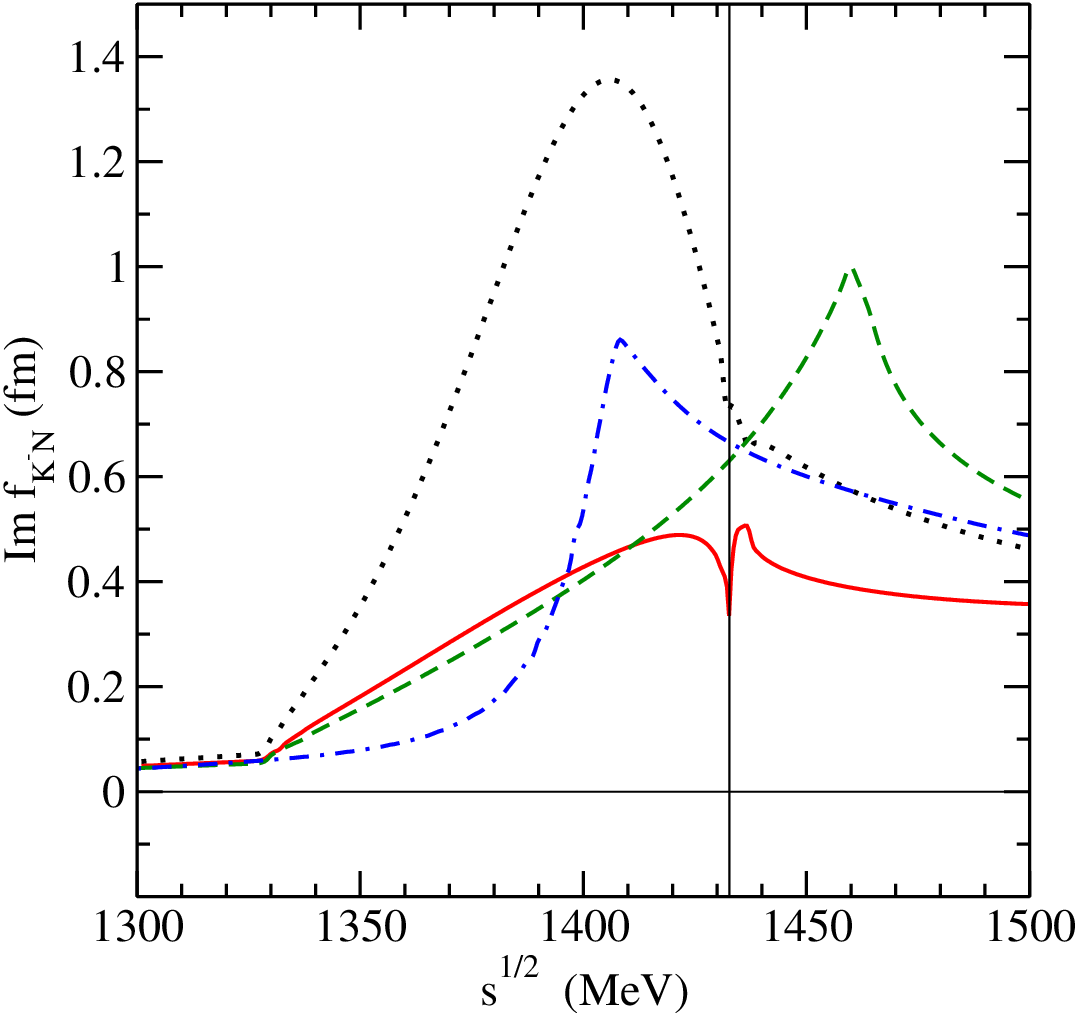}
\end{center}
\caption{Energy dependence of free-space (dotted line) amplitude $f_{K^- N}=\frac{1}{2}(f_{K^- p}+f_{K^- n})$ 
compared with WRW modified amplitude (solid line), Pauli (dashed line), and Pauli + SE (dot-dashed line) modified amplitude for $\rho_0=0.17$~fm$^{-3}$ in the P model (left: real parts, right: imaginary parts).The thin vertical line indicates the $K^-N$ threshold. }
\label{fig.:wrw_pauli}
\end{figure}

In previous calculations~\cite{cfggmPLB, gmNPA}, the in-medium modifications of the $K^- N$ amplitudes in the P model~\cite{pnlo} were accounted for in a 
different way. The integration over the intermediate meson-baryon momenta in the underlying Green's function was restricted to a region 
ensuring the nucleon intermediate energy to be above the Fermi level (denoted further `Pauli'). Moreover, the in-medium hadron self-energies 
(denoted `Pauli+SE') were considered in some cases as well. In Fig.~\ref{fig.:wrw_pauli}, we compare the Pauli correlated amplitudes with the WRW 
modified amplitudes in the P model. Both approaches, WRW and Pauli, yield similar $K^-N$ in-medium reduced amplitudes\footnote{$F_{K^-N} = g(p)f_{K^-N}g(p')$, where $g(p)$ is a momentum-space form factor (see Ref.~\cite{cfggmPLB})} $f_{{K^-}N}=\frac{1}{2}(f_{K^- p}+f_{K^- n})$ 
in the subthreshold energy region. Above threshold, the behavior of Pauli and WRW modified amplitudes is different. 
The effect of hadron self-energies is illustrated in Fig.~\ref{fig.:wrw_pauli} as well. The Pauli correlated and Pauli+SE amplitudes are again quite similar to each other 
farther below threshold (in the region relevant to $K^-$-nuclear bound state calculations), but they differ appreciably near and above threshold. 

The existence of the subthreshold resonance $\Lambda(1405)$, which is dynamically generated in chirally-motivated coupled-channel models, causes that the $K^- p$ amplitudes exhibit strong energy (and density) dependence near and below threshold. This feature requires a proper self-consistent scheme for evaluating the $K^-$ optical potential in both calculations of $K^-$ atomic as well as nuclear states~\cite{cfggmPLB, gmNPA, fgNPA, fgNPA16}.\\ 
The in-medium amplitudes entering Eq.~\eqref{Eq.:in-med amp} are a function of energy $\sqrt{s}$ given by Mandelstam variable 
\begin{equation}
 s=(E_N+E_{K^-})^2-(\vec{p}_N+\vec{p}_{K^-})^2~,
\end{equation}
where $E_N=m_N-B_N$, $E_{K^-}=m_{K^-}-B_{K^-}-V_C$ and $\vec{p}_{N(K^-)}$ is the nucleon (kaon) momentum. Unlike the free two-body cm system, the momentum dependent term $(\vec{p}_N+\vec{p}_{K^-})^2 
\neq 0$ in the $K^-$-nucleus cm frame, which generates additional substantial downward energy 
shift \cite{cfggmPLB}. The non-negligible momentum term is upon averaging over angles equal to 
$p^2_{K^-} + p^2_N$. This averaging, i.e. dropping the term $\sim \vec{p_{K^-}} \cdot \vec{p_N}$, has been meant to provide a mean value of the energy $\sqrt{s}$ for a given density. It is not a substitute for a proper treatment of Fermi motion. The effect of Fermi motion was studied in detail in Ref.~\cite{wkwPLB96} where it was demonstrated that the Fermi averaging has a small effect on the $K^-$ binding energy. Nevertheless, we performed calculations using averaging on the level of $K^-N$ amplitudes instead of angular averaging. We verified that both approaches yield very similar results --- $K^-$ binding energies differ by $\leq 2$\% and the widths by $\leq 10$\%. 

The kaon kinetic energy is given in the local density approximation by 
\begin{equation} \label{kaon_kin_en}
 \frac{p^2_{K^-}}{2m_{K^-}}= -B_{K^-} - \text{Re}V_{K^-} - V_C~,
\end{equation}
where $V_{K^-}$ is the $K^-$-nuclear optical potential.  
The nucleon kinetic energy is expressed within the Fermi gas model as 
\begin{equation}
 \frac{p_N^2}{2m_N}=T_N \left(\frac{\rho}{\bar{\rho}}\right)^{2/3}~,
\end{equation}
where $T_N=23$~MeV is the average nucleon kinetic energy and $\bar{\rho}$ is the average nuclear density distribution.

Finally, the $K^-N$ amplitudes can be expressed as a function of energy $ \sqrt{s} = E_{\rm th} + \delta \sqrt{s}$ where $E_{\rm th}=m_N + m_{K^-}$ and the energy shift $\delta \sqrt{s}$ is expanded near threshold in terms of binding and 
kinetic energies (to leading order):
\begin{equation}\label{Eq.:deltaEs}
 \delta \sqrt{s} \approx  -B_N - B_{K^-} -V_C -\beta_N T_N\left(\frac{\rho}{\bar{\rho}}\right)^{2/3} - \beta_{K^-}\left( -B_{K^-} - {\rm Re}V_{K^-}(r) -V_C \right)
\end{equation}
where $\beta_{N(K^-)}={m_{N(K^-)}}/(m_N+m_{K^-})$ and $B_N=8.5$~MeV is the average binding energy per nucleon.
After introducing specific forms of density dependence ensuring that $\delta \sqrt{s} \rightarrow 0$ as $\rho \rightarrow 0$ in agreement with the low-density limit (for details see Ref.~\cite{fgNPA}) 
the energy shift $\delta \sqrt{s}$ in Eq.~\eqref{Eq.:deltaEs} has the following form:
\begin{equation} \label{Eq.:deltaEsLDL}
 \delta \sqrt{s}=  -B_N\frac{\rho}{\bar{\rho}}\, - \beta_N\! \left[B_{K^-}\frac{\rho}{\rho_{\rm max}} + T_N\left(\frac{\rho}{\bar{\rho}}\right)^{2/3}\!\!\!\! +V_C\left(\frac{\rho}{\rho_{\rm max}}\right)^{1/3}\right] + \beta_{K^-} {\rm Re}V_{K^-}(r)~,
\end{equation}
where $\rho_{\rm max}$ is the maximal value of the nuclear density. The $K^-$ binding energy $B_{K^-}$ is multiplied by $\rho/\rho_{\rm max}$, which ensures that the $K^-$ kinetic energy expressed in Eq.~\eqref{kaon_kin_en} in terms of local density approximation is positive at any nuclear density.

It is to be noted that since the input of our work was adopted from the kaonic atoms analysis of Friedman and Gal \cite{fgNPA16}, it is desirable to keep consistent and use similar kinematics in our calculations.

\begin{figure}[t]
\begin{center}
\includegraphics[width=0.5\textwidth]{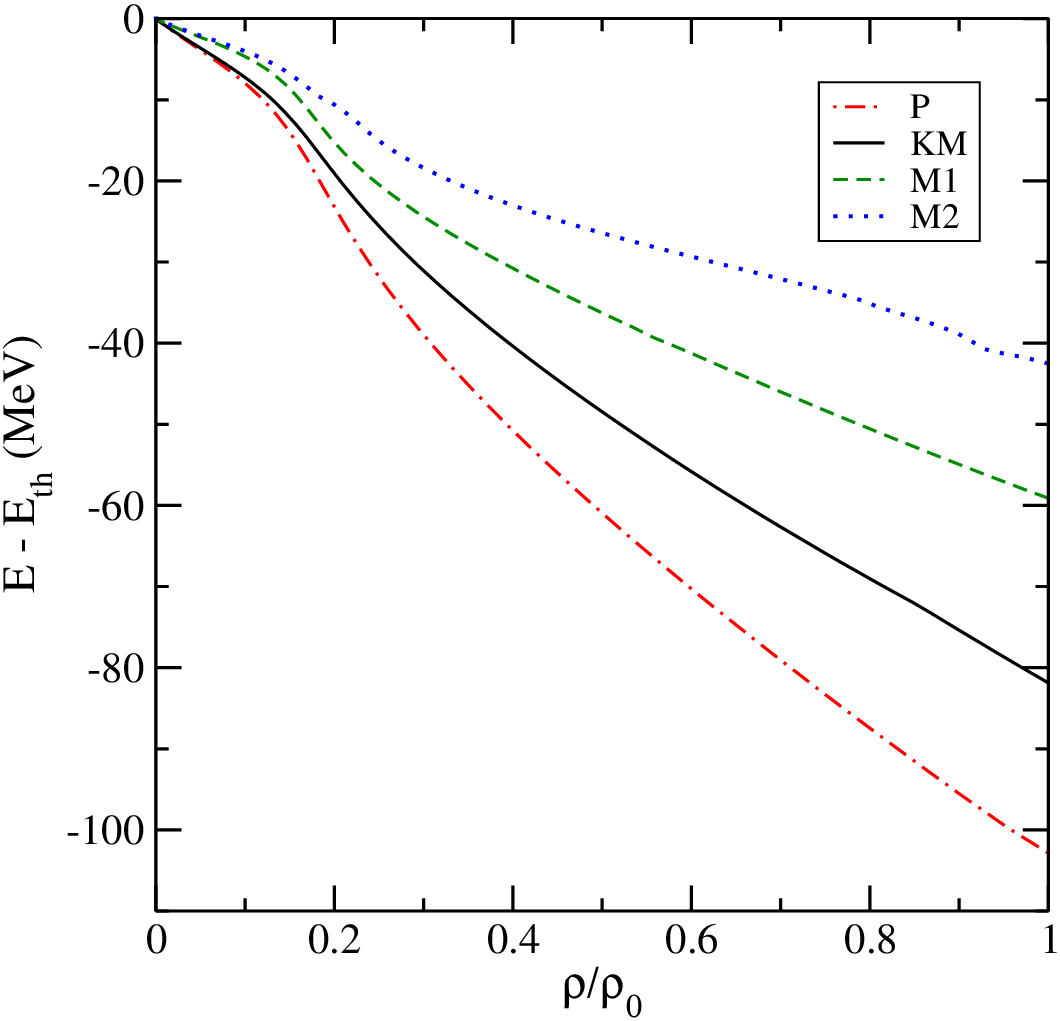} 
\end{center}
\caption{Subthreshold energies probed in the $^{16}$O$+K^-$ nucleus 
as a function of relative density $\rho/ \rho_0$, calculated self-consistently using  $K^-N$ amplitudes in the P (dot-dashed line), KM (solid line), M1 (dashed line), and M2 (dotted line) models.}
\label{fig.:deltaEs}
\end{figure}

In Fig.~\ref{fig.:deltaEs} we present the downward energy shift $\delta \sqrt{s} = E - E_{\rm th}$ 
as a function of relative density $\rho / \rho_0$ probed in the self-consistent calculations 
with in-medium $K^-$ optical potential $V_{K^-}^{(1)}$ based on amplitudes from chiral models P, 
KM, M1, and M2. The calculations were performed for the  $^{16}$O+$K^-$ system. 
The models considered here predict quite different energy shifts, reaching at the saturation density values between $\sim - 40$~MeV for the M2 model and $\sim  - 100$~MeV for the P model.
The energy shifts corresponding to the Bonn models B2 and B4 are not plotted in the figure since 
these models do not yield any $K^-$-nuclear bound state. It is to be noted that though the free-space amplitudes in Fig.~\ref{fig.:KpnFree} are shown only to $\sqrt{s}=1370$~MeV, the amplitudes for KM and P models are available down to $1300$~MeV. The energy shifts $\delta \sqrt{s}$ in the models shown in Fig.~\ref{fig.:deltaEs} are thus safely in the available energy region.

\begin{figure}[t]
\begin{center}
\includegraphics[width=0.9\textwidth]{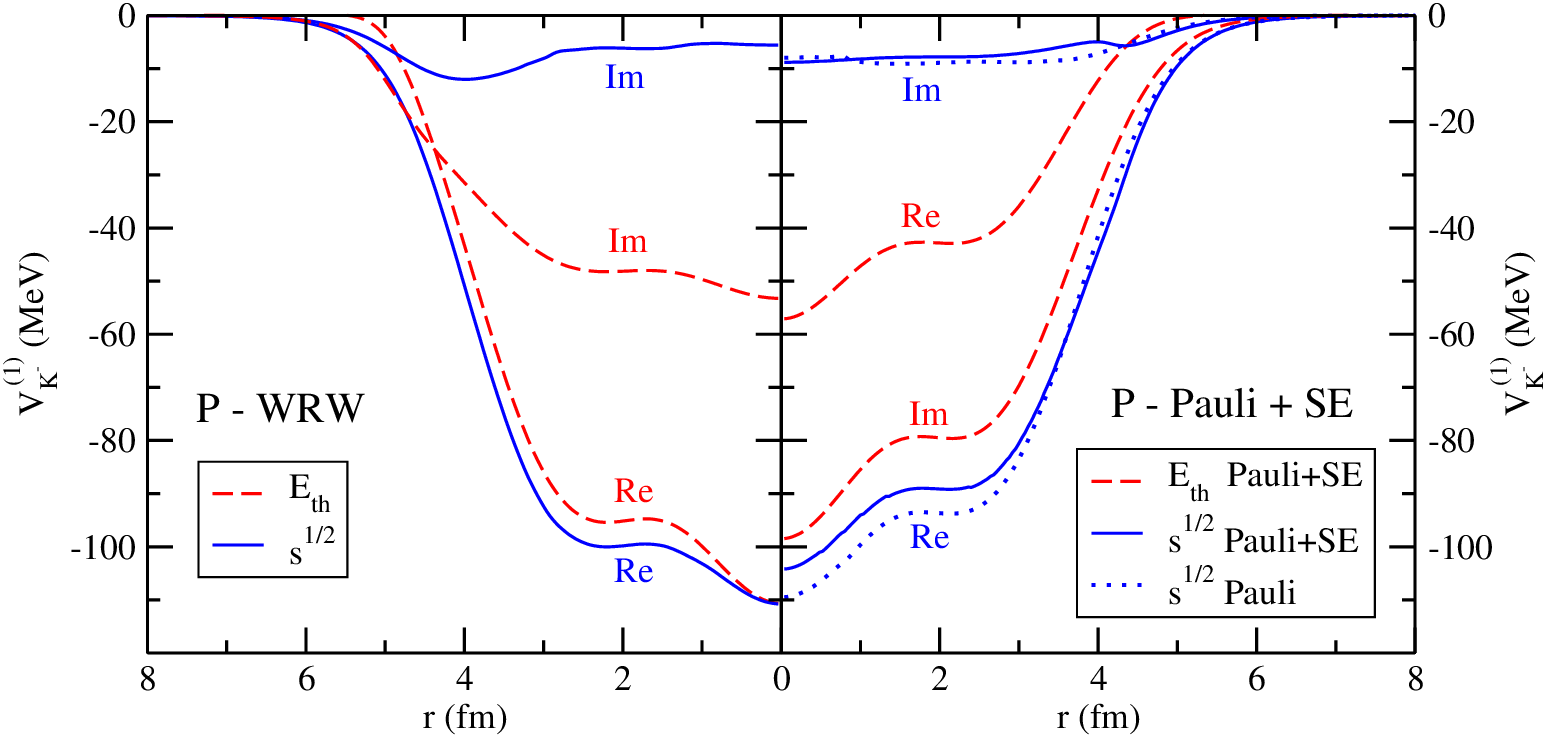}
\end{center}
\caption{$K^-$ nuclear potential $V_{K^-}^{(1)}$ in $^{40}$Ca calculated using chiral $K^- N$ P amplitudes at threshold (dashed lines) and with $\sqrt{s}$ (Eq.~(8) of Ref.~\cite{gmNPA}) (solid lines), in two in-medium versions: WRW (left panel) and Pauli+SE (right panel). The Pauli version (right panel, dotted line) for $\sqrt{s}$ from \cite{gmNPA} is shown as well (see text for details).}
\label{fig.:pnlo+se}
\end{figure} 

In calculations presented in this work, we take into account only Pauli correlations in the medium expressed within the WRW approach. One might argue that the effect of hadron self-energies should be included as well. 
In Fig.~\ref{fig.:pnlo+se} we demonstrate the role of hadron self-energies in $^{40}$Ca. 
We compare the $K^-$ potential $V_{K^-}^{(1)}$ calculated in the P model within the WRW method (left panel) with the $K^-$ potential 
calculated using the Pauli and Pauli +SE in-medium amplitudes, used in previous calculations of $K^-$-nuclear 
bound states \cite{gmNPA} (right panel). 
The hadron self-energies modify considerably the potential evaluated at threshold while 
their effect becomes rather small in self-consistent treatment of the energy shift. Then the WRW, Pauli and Pauli+SE options for in-medium modifications of $K^- N$ amplitudes give nearly identical $K^-$-nucleus potentials. 


\begin{figure}[t]
\begin{center}
\includegraphics[width=0.45\textwidth]{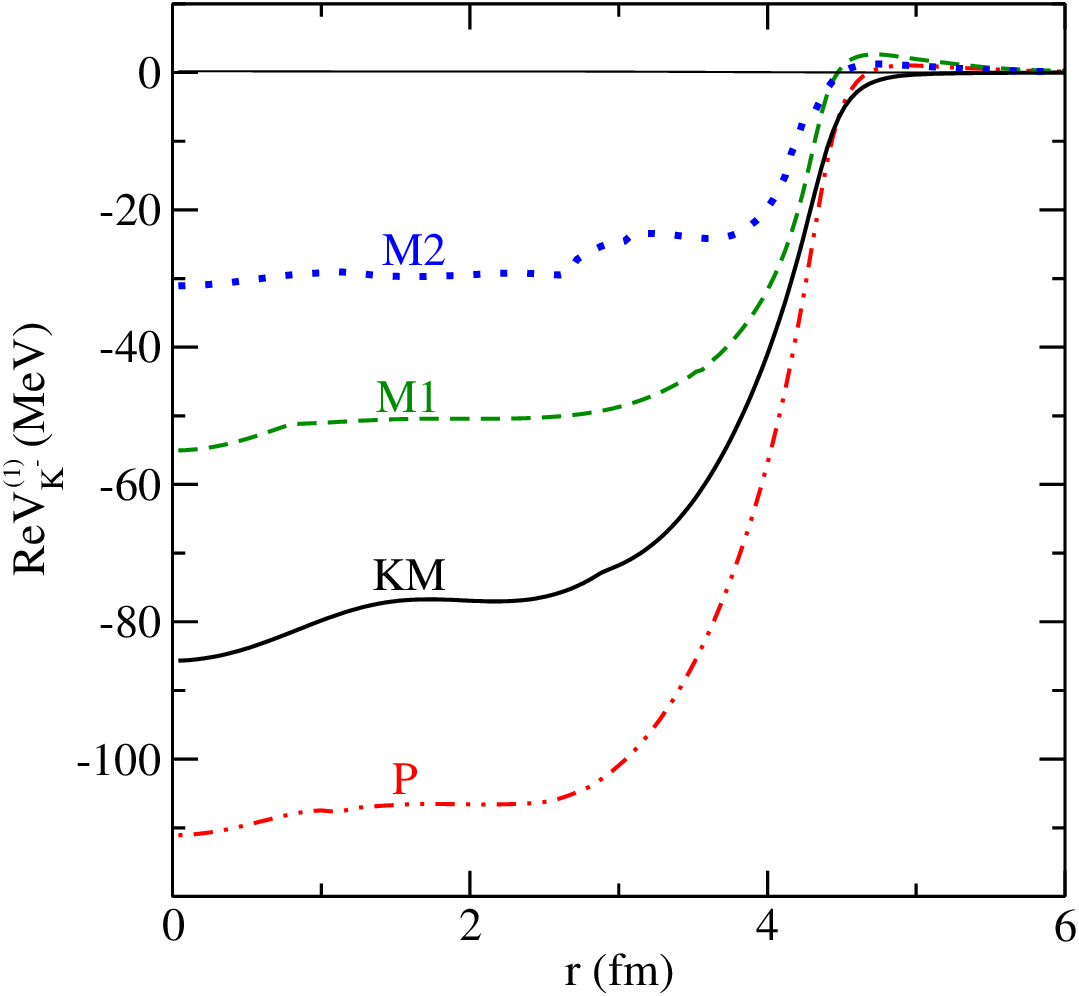} \hspace{20pt}
\includegraphics[width=0.45\textwidth]{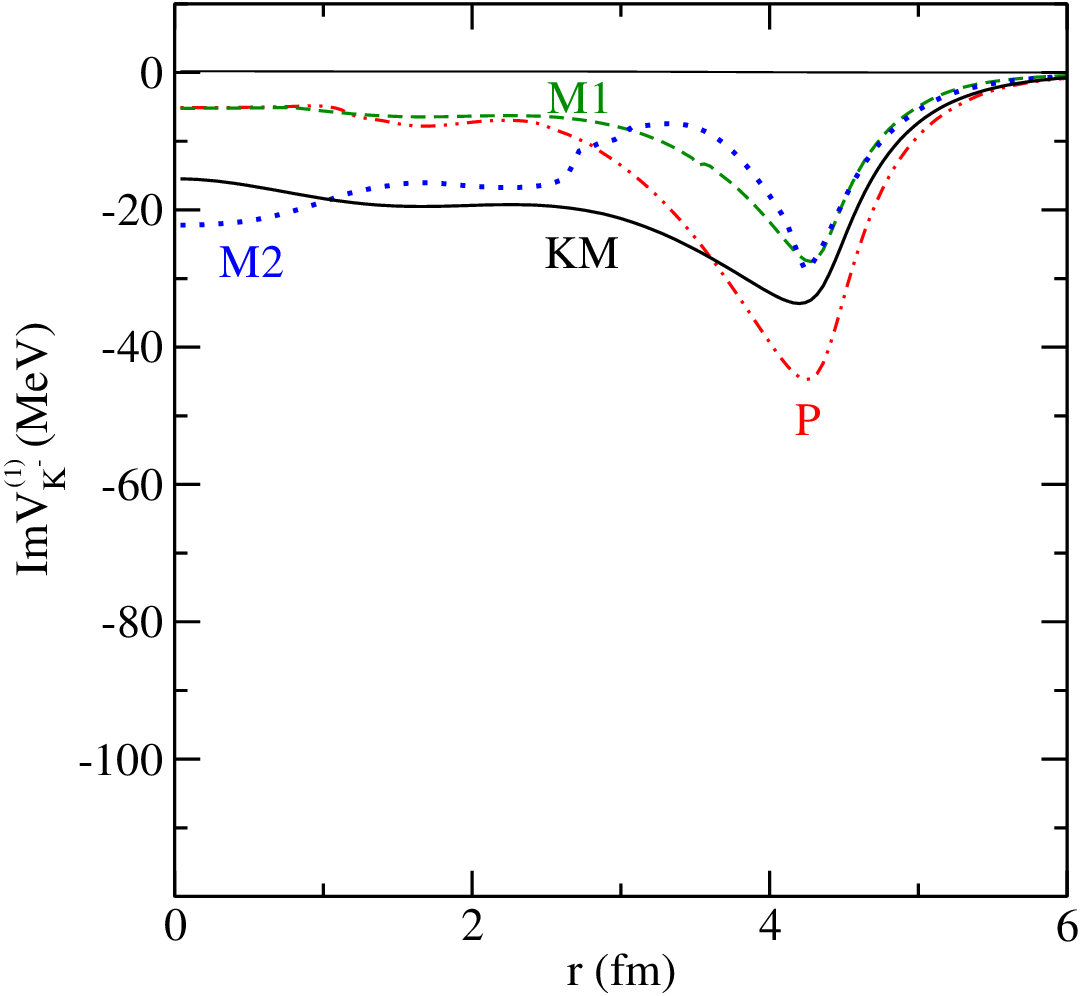}
\end{center}
\caption{Real (left) and imaginary (right) parts of the $K^-$ nuclear potential $V_{K^-}^{(1)}$ in $^{40}$Ca calculated self-consistently using chiral P (dot-dashed line), KM (solid line), M1 (dashed line), and M2 (dotted line) amplitudes.}
\label{fig.:potcapkmm1}
\end{figure}
As was shown in Figs.~1 and 2, the chiral $K^- N$ amplitudes differ considerably below threshold, thus in the region relevant 
to calculations of kaonic nuclear states. As a consequence, corresponding $K^-$-nucleus potentials derived using these 
amplitudes differ significantly as well. In Fig.~\ref{fig.:potcapkmm1}, we present real (left) and imaginary (right) parts of 
the $K^-$-nuclear optical potential $V_{K^-}^{(1)}$ in $^{40}$Ca, calculated self-consistently within P, KM, M1, and M2 models. 
The depths of Re$V_{K^-}^{(1)}$ are ranging from 30~MeV in the M2 model to 110~MeV in the P model. 
The imaginary parts of the $K^-$ potentials are rather shallow inside the nucleus, which reflects sizable downward energy shift 
to the vicinity of threshold of the main decay channel $K^- N \rightarrow \pi\Sigma$. The apparent dip in the surface region 
is due to the low-density limit adopted in $\delta \sqrt{s}$ (see Eq.~\eqref{Eq.:deltaEsLDL}). 

\begin{figure}[t!]
\begin{center}
\includegraphics[width=0.45\textwidth]{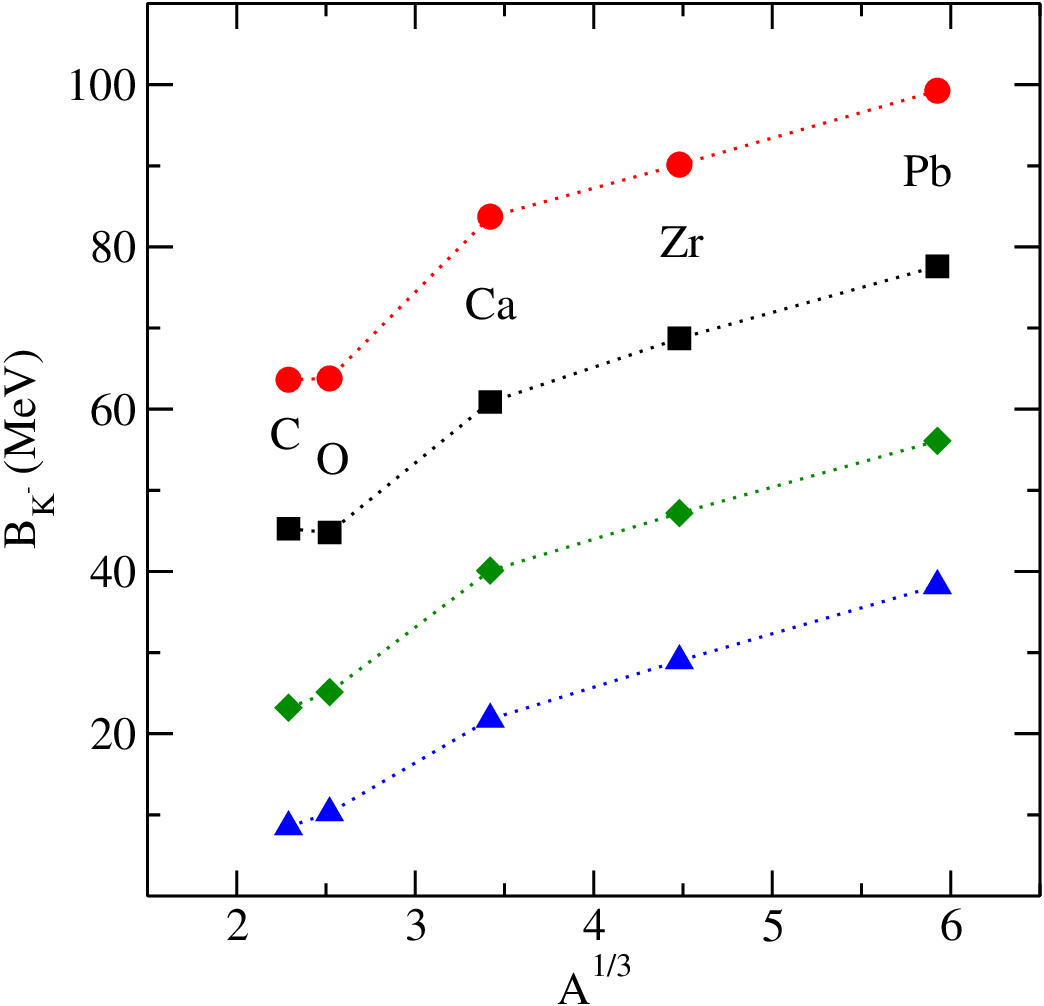} \hspace{20pt}
\includegraphics[width=0.45\textwidth]{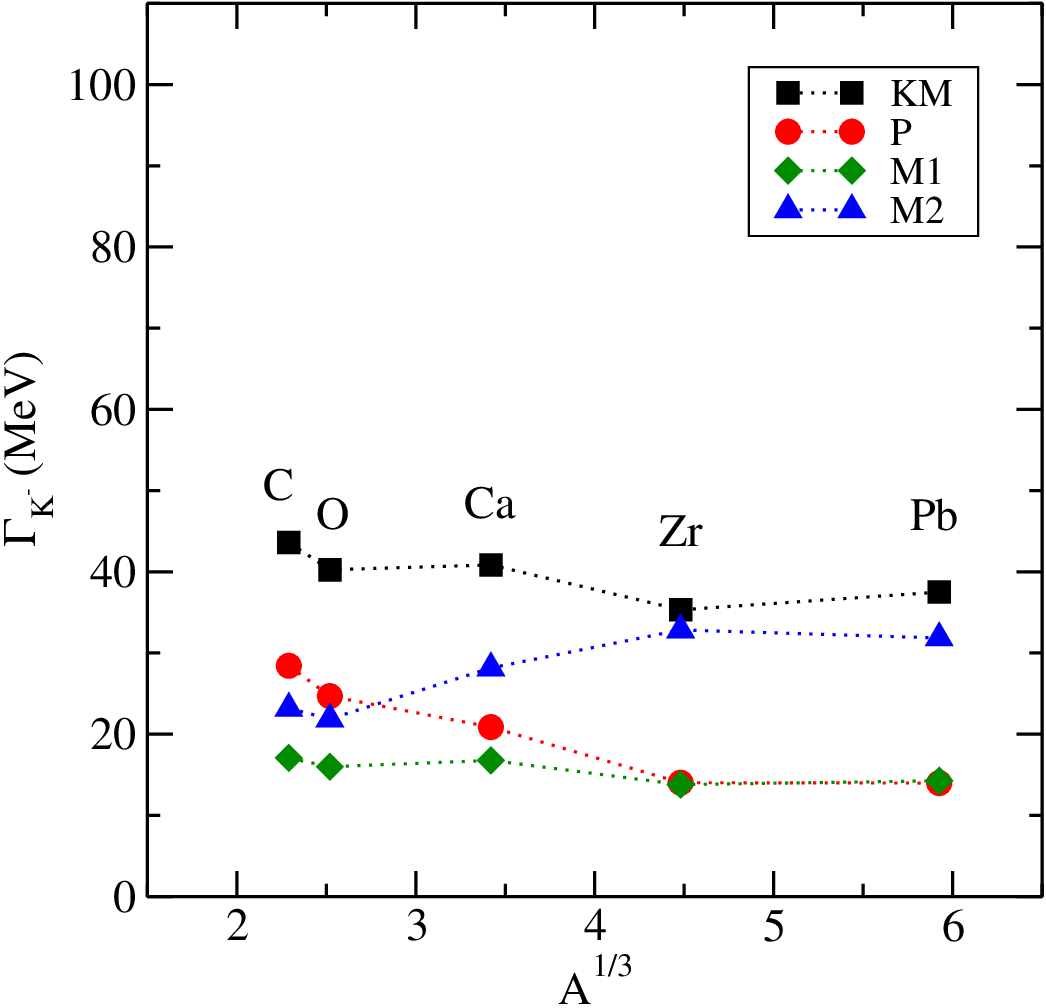}
\end{center}
\caption{1s $K^-$ binding energies (left) and corresponding widths (right) in various nuclei calculated self-consistently in P (circles), KM (squares), M1 (diamonds), and M2 (triangles) models. 
$K^-$-multinucleon interactions are not considered.}
\label{BkGk}
\end{figure}

The $1s$ binding energies $B_{K^-}$ and widths $\Gamma_{K^-}$ in selected nuclei are presented in Fig.~\ref{BkGk}. 
The calculated $K^-$ binding energies are strongly model dependent due to different depths of Re$V_{K^-}^{(1)}$ in various $K^- N$ 
interaction models. However, they exhibit similar $A$ dependence in all models considered. The $K^-$ widths are rather small 
and weakly $A$-dependent. The KM model predicts widths up to three times larger than the P and M1 models. The M2 model yields similar 
widths as the KM model for $^{208}$Pb and $^{90}$Zr, while the widths in lighter nuclei are comparable with the P model widths. 
It is to be noted that we get no kaonic nuclear bound states for the Bonn models B2 and B4 because the real parts of the 
in-medium $K^-N$ amplitudes are repulsive in the relevant subthreshold region (see Figs.~1 and 2). 

\begin{figure}[h!]
\begin{center}
\includegraphics[width=0.45\textwidth]{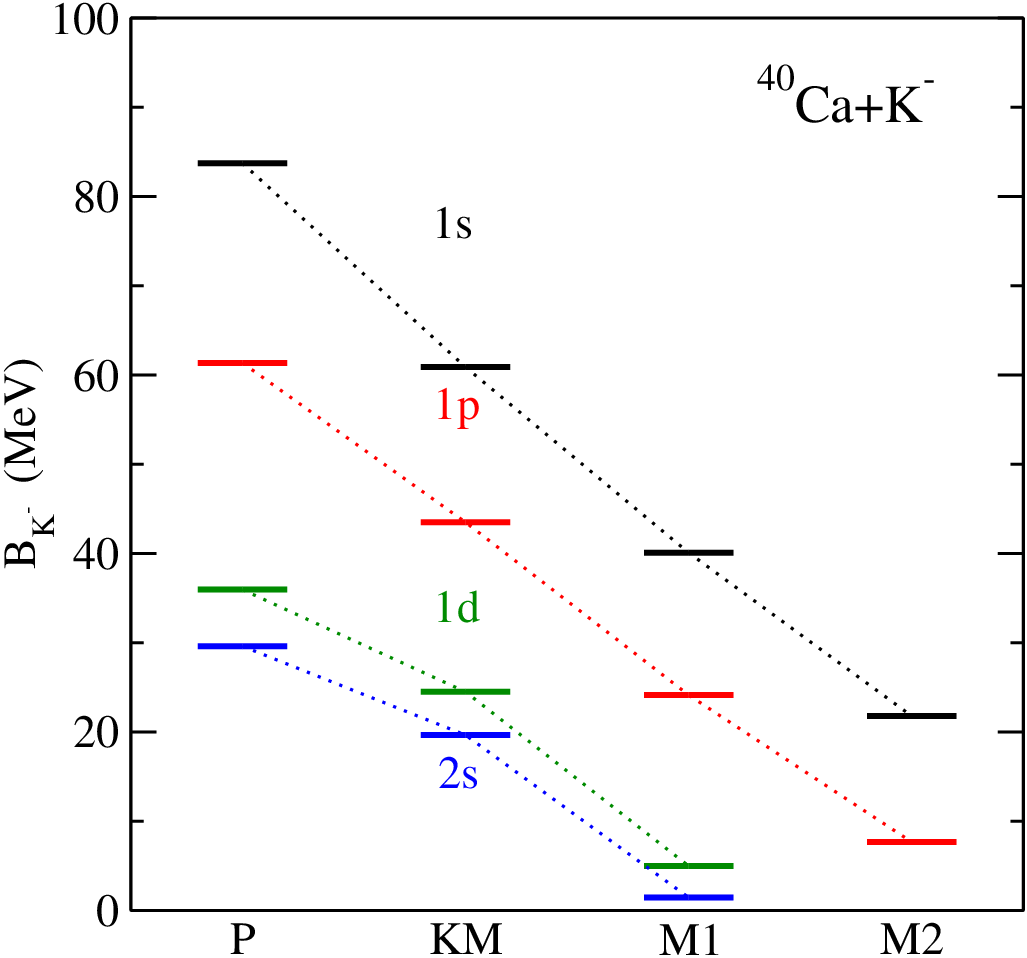} \hspace{20pt}
\includegraphics[width=0.45\textwidth]{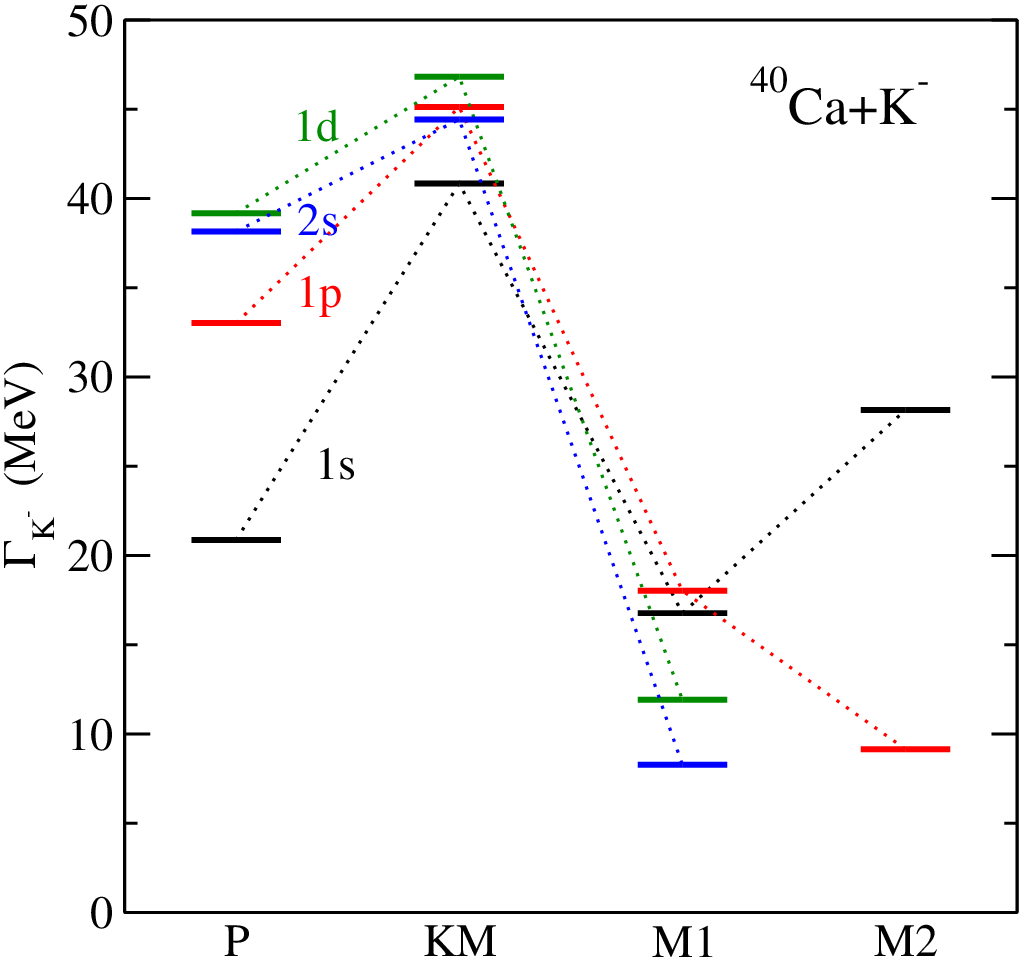}
\end{center}
\caption{$K^-$ binding energies (left) and widths (right) in $s, p$ and $d$ levels in $^{40}$Ca calculated self-consistently in P, KM, M1, and M2 models. $K^-$-multinucleon interactions are not 
considered.}
\label{BkGkCa}
\end{figure}

In Fig.~\ref{BkGkCa} (left panel) we compare $K^-$-nuclear single-particle spectra in $^{40}$Ca, calculated using various $K^- N$ 
interaction models. Again, the $K^-$ binding energies $B_{K^-}$ strongly depend on the model used.  
The relative position of the $K^-$ spectra is in accordance with the depths of the $K^-$-nucleus potentials $V_{K^-}^{(1)}$ shown in 
Fig.~6.   
The corresponding $K^- N \rightarrow \pi Y$ conversion widths are presented in Fig.~\ref{BkGkCa} 
(right panel). In the P and KM models, the $1s$-state widths are reduced due to considerable 
energy shift towards the $\pi \Sigma$ threshold and become smaller than the widths of excited states, for which  
$\sqrt{s}$ is farther from the $\pi \Sigma$ threshold. On the other hand, the $K^-$ widths calculated in M1 and M2 
models follow the opposite trend. It is because $\sqrt{s}$ in these models is much closer to the $K^-N$ threshold 
where the (dominant) imaginary part of the $K^-p$ amplitudes starts to decrease towards the threshold (see Fig.~2). 
This feature is more pronounced in the M2 model which gives a smaller downward energy shift due to the shallower 
$K^-$ potential ($1d$ and $2s$ states are unbound).  
\\

Following results of calculations presented so far, one might conclude that at least some $K^- N$ interaction models predict 
sufficiently bound kaonic nuclear states with relatively narrow widths. In the nuclear medium, however,  $K^-$ multinucleon processes take place as 
well. They are becoming more and more important with increasing nuclear density and $K^-$ binding energy \cite{mfgPLB, fgmNPA}. We will demonstrate their significant role 
in self-consistent calculations of kaonic nuclei in the next Section. 

\section{The role of $K^-$ multinucleon interactions}
\label{sec-2}
The $K^-$ multinucleon interactions are an inseparable component of every realistic description of $K^-$-nucleus interaction. As was shown in recent analysis by Friedman and Gal \cite{fgNPA16}, the single-nucleon $K^-$ potential constructed within all chiral meson-baryon interaction models considered in this work has to be supplemented by a phenomenological term representing $K^-$ multinucleon processes in order to obtain good fit to kaonic atom data. The total $K^-$ optical potential is then a sum of single-nucleon and multinucleon potential $V_{K^-}=V_{K^-}^{(1)}+V_{K^-}^{(2)}$, where the single-nucleon potential $V_{K^-}^{(1)}$ is given by Eq.~\eqref{piK} and the multinucleon term $V_{K^-}^{(2)}$ is of the form
\begin{equation} \label{Vknn}
 2\text{Re}(\omega_{K^-})V_{K^-}^{(2)}=-4 \pi B (\frac{\rho}{\rho_0})^{\alpha} \rho~.
\end{equation}
The values of the complex amplitude $B$ and positive exponent
$\alpha$ listed in Table~\ref{Tab.:ampB} were obtained by fitting kaonic atom data for each $K^-N$ amplitude model separately~\cite{fgNPA16}. Moreover, the total $K^-$ optical potentials $V_{K^-}$ were then confronted with branching ratios of $K^-$ absorption at rest. Only two models, P and KM, were found to reproduce simultaneously the fractions of $K^-$ single-nucleon absorption from bubble chamber experiments \cite{bubble1, bubble2, bubble3} and kaonic atom data. Yet, we performed calculations 
for all six discussed $K^- N$ amplitude models. It is to be noted that the P and KM models could be regarded as equivalent within the uncertainties shown in Table~\ref{Tab.:ampB}.

\begin{table}[t]
\begin{center}
\caption{Values of the complex amplitude $B$ and exponent $\alpha$ used to evaluate $V_{K^-}^{(2)}$  for all chiral 
meson-baryon interaction models considered in this work.}\vspace{3pt}
 \begin{tabular}{c|cccc}
  & P1  & KM1 & P2 & KM2  \\ \hline 
$\alpha$  &  1 & 1 & 2 & 2\\ 
Re$B$ (fm) & -1.3 $\pm$ 0.2 & -0.9 $\pm$ 0.2 & -0.5 $\pm$ 0.6 & 0.3 $\pm$ 0.7 \\ 
Im$B$ (fm) & ~1.5 $\pm$ 0.2 & ~1.4 $\pm$ 0.2 & ~4.6 $\pm$ 0.7 & 3.8 $\pm$ 0.7\\ \hline
 & & & & \\[-20pt]
  & B2 & B4 & M1 & M2  \\ \hline 
$\alpha$  & 0.3 & 0.3 & 0.3 & 1 \\ 
Re$B$ (fm) & 2.4 $\pm$ 0.2 & 3.1 $\pm$ 0.1 & 0.3 $\pm$ 0.1 & 2.1 $\pm$ 0.2 \\ 
Im$B$ (fm) & 0.8 $\pm$ 0.1 & 0.8 $\pm$ 0.1 & 0.8 $\pm$ 0.1 & 1.2 $\pm$ 0.2 
 \end{tabular}
 \label{Tab.:ampB}
\end{center}
\end{table}

The dominant mode of $K^-$ absorption on two nucleons in the nuclear interior is the non-pionic conversion $K^-NN \rightarrow \Sigma N$ \cite{sjPRC12, mfgPLB, sjPRC79}. Since the amplitude Im$B$ is constant, we multiply it by kinematical suppression factor to account for phase space reduction for decay products in $K^-NN\rightarrow \Sigma N$ absorption in 
the nuclear medium. The suppression factor used in our calculation is of the form
\begin{equation}
f_{\Sigma N}=\frac{M^3}{s_{m}^{3/2}}\sqrt{\frac{[s_m-(m_N+m_{\Sigma})^2][s_m-(m_N-m_{\Sigma})^2]}{[M^2-(m_N+m_{\Sigma})^2][M^2-(m_N-m_{\Sigma})^2]}}\Theta(\sqrt{s}_m-m_N-m_{\Sigma})~,
\end{equation}
where $M=2m_N+m_{K^-}$ and $\sqrt{s}_m=M-\delta \sqrt{s}$ \cite{mfgPLB}. \\  
It is to be noted that for processes on a single nucleon, the proper energy dependence is embedded 
directly in the $K^-N$ amplitudes constructed within chirally-motivated coupled-channel models.

Analyses of Friedman and Gal have shown that kaonic atom data constrain reliably the real part of the $K^-$ optical potential up to $\sim 25$\% of $\rho_0$ and its imaginary part up to $\sim 50$\% of $\rho_0$. The shape of the phenomenological $K^-$ optical potential $V_{K^-}^{(2)}$ in the nuclear interior is thus a matter of extrapolation to higher densities. In order to allow for more flexibility, we consider different options for $V_{K^-}^{(2)}$  
beyond the half density limit $\rho(r) = 0.5 \rho_0$ in our calculations. 
First, the form~\eqref{Vknn} is applied in the entire nucleus 
(full density option - FD). Second, the potential $V_{K^-}^{(2)}$ is fixed at constant value $V_{K^-}^{(2)}(0.5\rho_0)$  
for $\rho (r) \ge 0.5 \rho_0$ (half density limit - HD). In the third approximation (TR), the $t\rho$ form of $V_{K^-}^{(2)}$ is assumed for densities $\rho(r) \ge 0.5\rho_0$ in Eq. \eqref{Vknn}, i.e. 
$V_{K^-}^{(2)} \sim -4 \pi B (0.5)^{\alpha} \rho$ for $\rho(r) \ge 0.5\rho_0$.
 
\begin{figure}[t!]
\begin{center}
\includegraphics[width=0.5\textwidth]{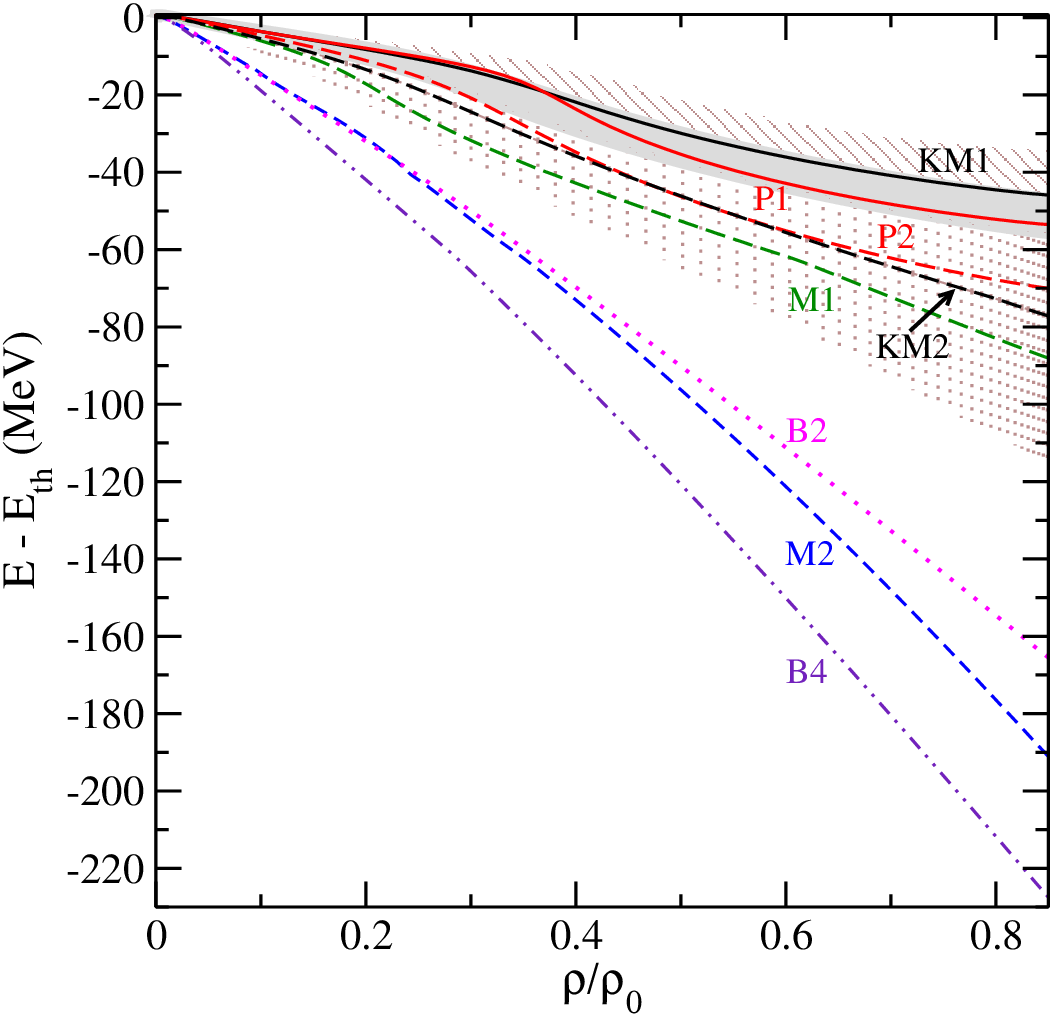}
\end{center}
\caption{
Subthreshold energies probed in the $^{208}$Pb$+K^-$ nucleus 
as a function of relative density $\rho/ \rho_0$, calculated self-consistently in all $K^-N$ amplitude models considered, supplemented by the FD variant of $V_{K^-}^{(2)}$. The dashed and dotted areas denote the uncertainty bands calculated in the KM1 and KM2 models and the shaded gray band represents their overlap.}
\label{rhodelta}
\end{figure}

\bigskip
In Fig.~\ref{rhodelta}, we present subthreshold energy shift $\delta\sqrt{s} = E - E_{\rm th}$  
as a function of the nuclear density in $^{208}$Pb, calculated in all $K^- N$ interaction models 
considered in this work, with the FD version of the $K^-$ multinucleon potential. For illustration, we show also the uncertainties involved in the 
KM1 and KM2 multinucleon potentials. They are denoted by dashed and dotted areas and the gray shaded band stands for their overlap. After including the $K^-$ multinucleon interactions 
in the KM and P models (the only two models accepted by analysis of Ref.~\cite{fgNPA16}), the energy shift 
$\delta \sqrt{s}$ for a particular 
density becomes smaller and moves back towards the $K^- N$ threshold (compare Fig.~\ref{fig.:deltaEs} and 
Fig.~\ref{rhodelta}). On the other hand, 
the B2, B4, and M2 models supplemented by a strongly attractive $K^-$ multinucleon potential 
Re$V_{K^-}^{(2)}$ (see Table~\ref{Tab.:ampB}) probe much deeper energy region below threshold than 
the KM and P models. In fact, fairly deep Re$V_{K^-}^{(2)}$, $(200 - 300)$~MeV, causes that $K^-$ 
will be bound even in the Bonn models B2 and B4. 

We witness large model dependence of the downward energy shifts $\delta\sqrt{s}$, ranging from -35 to -230~MeV  
in the nuclear center. This suggests that the models yield considerably different $K^-$ optical potentials. 
Yet, the KM and P models could be regarded as equivalent since they all lie in corresponding uncertainty 
bands and describe kaonic atom data equally well. We note that the free-space amplitudes in the M1, M2, B2, and B4 models were available only for  
$\sqrt{s}\geq 1370$~MeV. Therefore, we fixed the $K^-N$ amplitudes at 
constant value $F_{K^-N}(1370)$ when $\sqrt{s}$ got below 1370~MeV in our self-consistent calculations. 

\begin{figure}[t!]
\begin{center}
\includegraphics[width=0.45\textwidth]{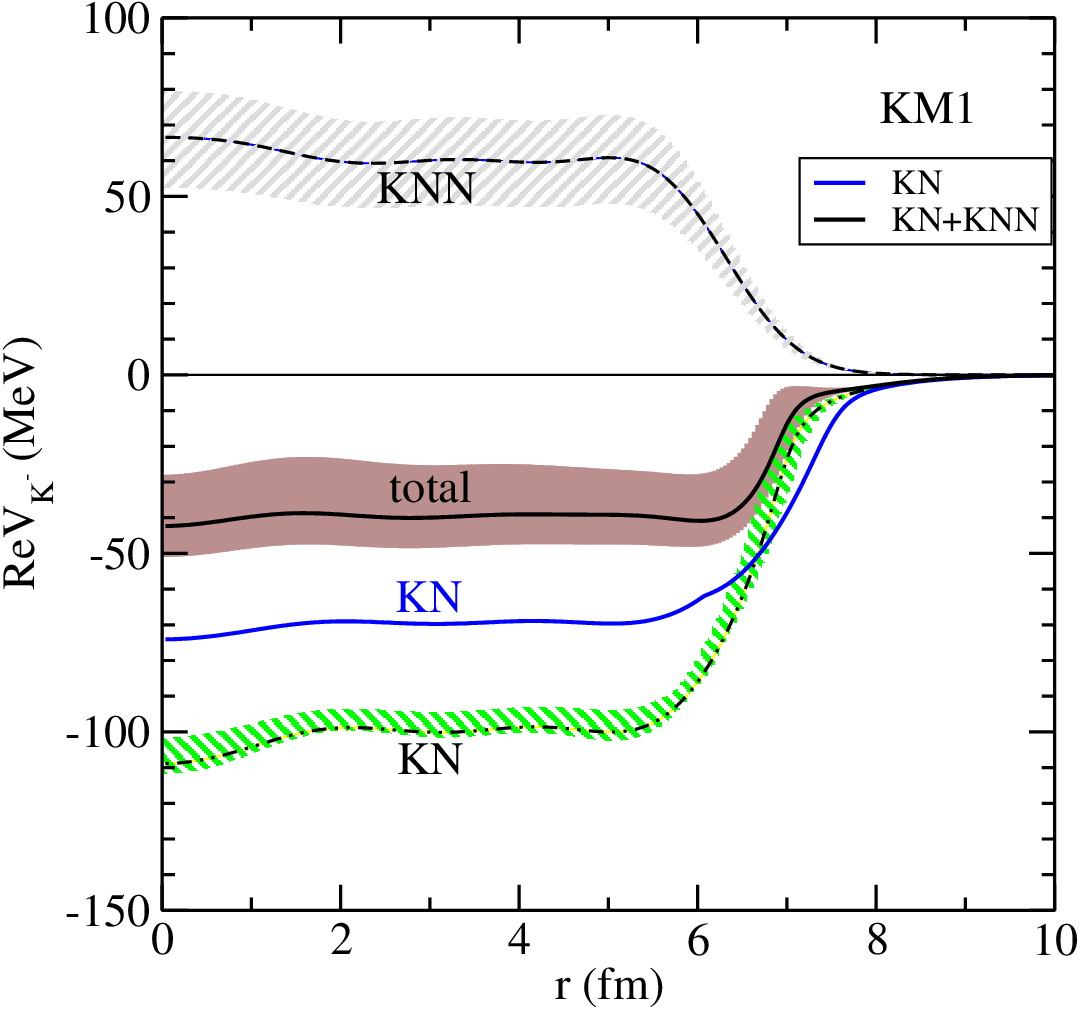} \hspace{20pt}
\includegraphics[width=0.45\textwidth]{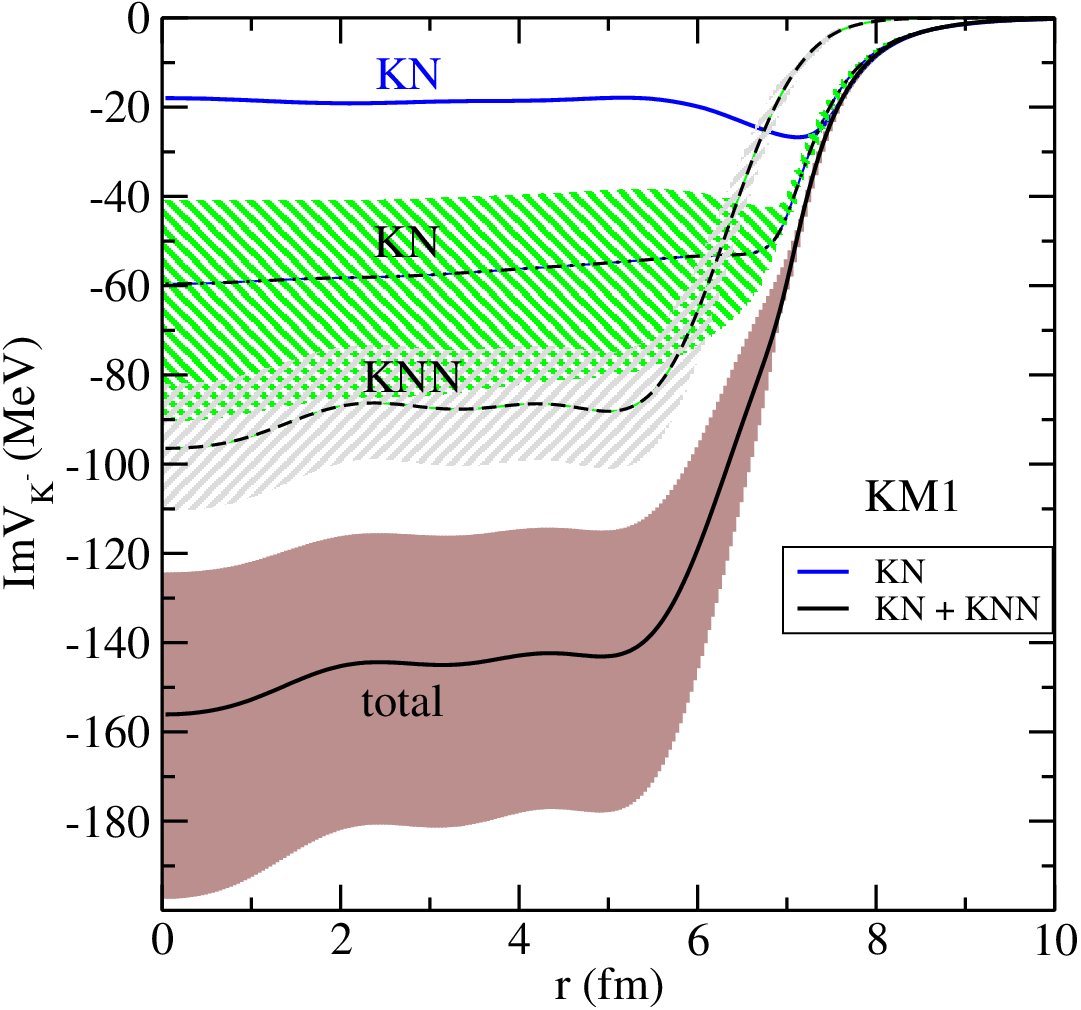} \\[20pt]
\includegraphics[width=0.45\textwidth]{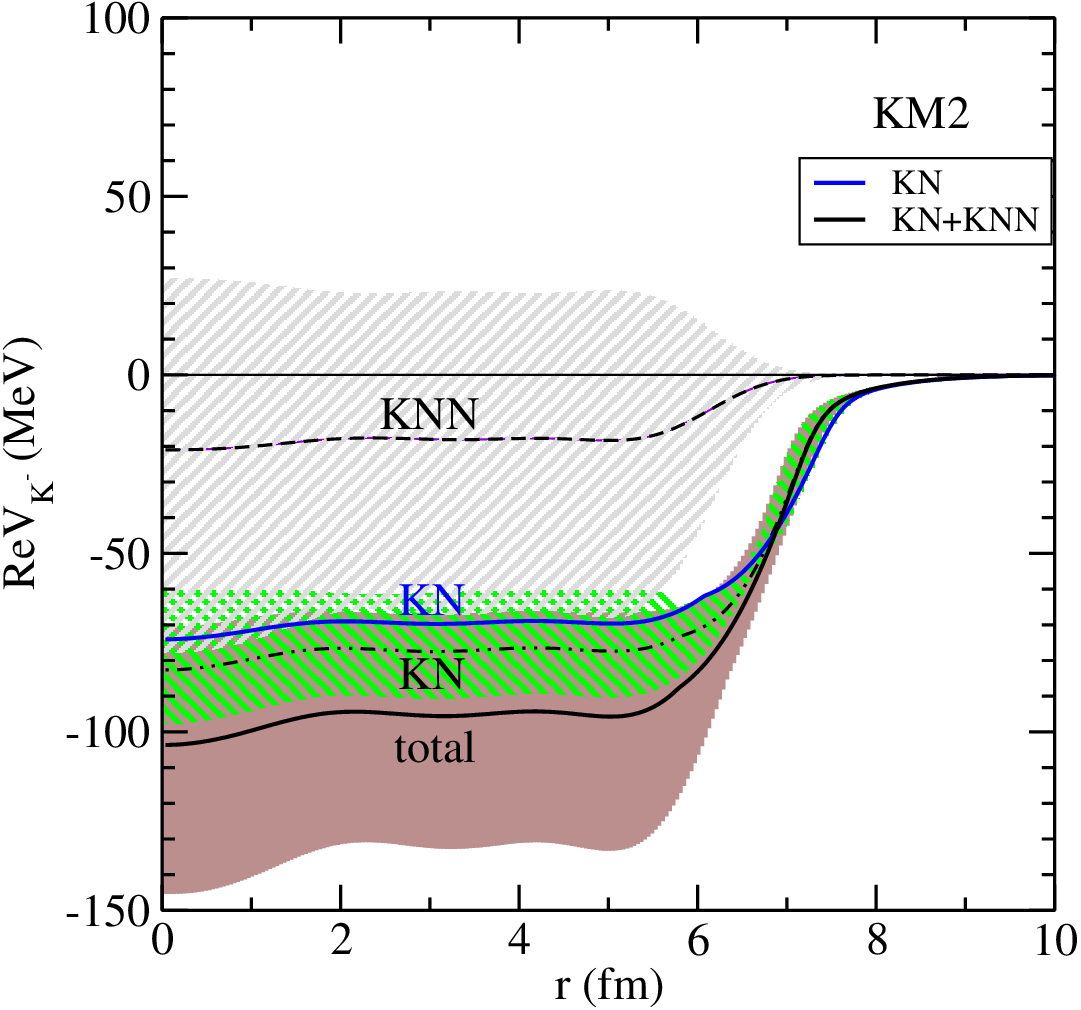} \hspace{20pt}
\includegraphics[width=0.45\textwidth]{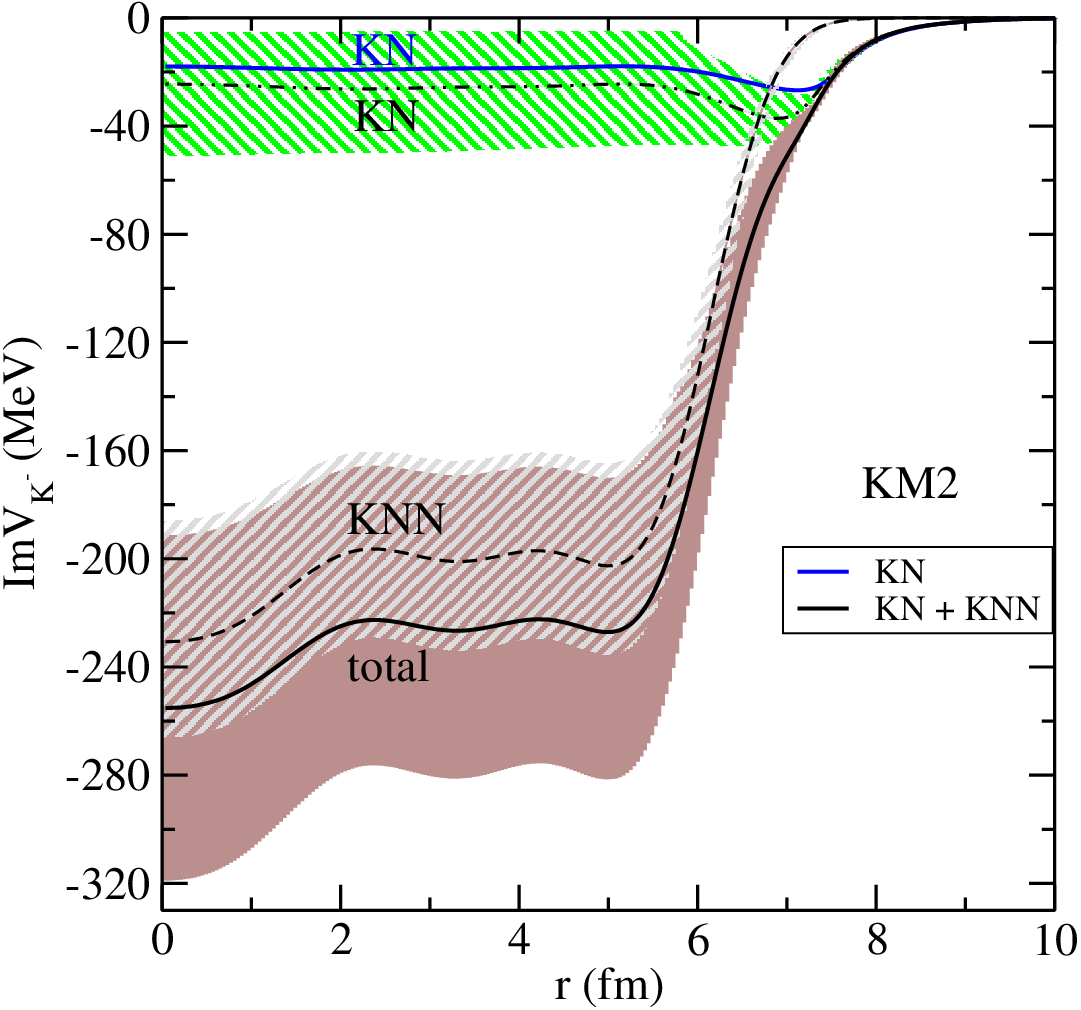}
\end{center}
\caption{The respective contributions from $K^-N$ (dashed dotted line) and $K^-NN$ (dashed line) 
potentials to the total real (left) and imaginary (right) $K^-$ optical potential in the 
$^{208}$Pb$+K^-$ nucleus, calculated self-consistently in the FD version of KM1 (top) and KM2 (bottom) models. The shaded areas denote the uncertainty bands. The $K^-$ single-nucleon potential (KN, blue solid line) calculated in the KM model (i.e. w/o multinucleon interactions) is shown for comparison.}
\label{potknn}
\end{figure}
The individual contributions from single-nucleon $V_{K^-}^{(1)}$ and multinucleon $V_{K^-}^{(2)}$ potentials to the total $K^-$ optical potential $V_{K^-}$ including their uncertainties (shaded areas) are shown in Fig.~\ref{potknn}, 
calculated self-consistently for $^{208}$Pb+$K^-$ in the KM1 (top panels) and  KM2 model (bottom panels) and the FD version of $V_{K^-}^{(2)}$.  For comparison, we present the single-nucleon $K^-N$ potential (KN, blue solid line) 
derived from the $K^-N$ amplitude model KM without considering multinucleon interactions.
The contribution from Re$V_{K^-}^{(2)}$ to the total real $K^-$-nucleus potential is repulsive in the KM1 model, 
as well as in the P1 and P2 models (not shown in the figure). As a result, the total $K^-$-nucleus potential including multinucleon 
processes is less attractive than the original single-nucleon $K^-$-nucleus potential. 
In the KM2 model the contribution from $V_{K^-}^{(2)}$ brings additional attraction to the total potential due to 
positive sign of the effective amplitude Re$B$ (see Table~\ref{Tab.:ampB}). However, 
the extensive uncertainty band in Fig.~\ref{potknn} proves that the sign of Re$B$ in the KM2 model is insignificant. 
The $V_{K^-}^{(1)}$ part of 
the optical potential in the KM1 and KM2  models (as well as in other models) differs from 
the original single-nucleon $K^-N$ potential due to the different subthreshold energy shift 
(see Fig.~\ref{rhodelta} and Fig.~\ref{fig.:deltaEs}). 
The uncertainties in the $K^-N$ part arise from variations of $\delta \sqrt{s}$ caused by the uncertainties in 
total $K^-$-nuclear potential. The depths of the total Re$V_{K^-}$ in the KM1(2), P1(2), and M1 models including 
the multinucleon potential $V_{K^-}^{(2)}$ are of the range $\simeq(50 - 100)$~MeV (not quoting uncertainties). \\ 

Adding $K^-$ multinucleon absorptions dramatically increases the depth of the total imaginary $K^-$ potential as 
illustrated in the right panels of Fig.~\ref{potknn}.
In the KM models (as well as P models, not shown in the figure), Im$V_{K^-}$ is much deeper than Re$V_{K^-}$ for both values of $\alpha$ even when the uncertainties 
are taken into account. The $K^-$ multinucleon processes contribute substantially to $K^-$ absorption mainly in 
the interior of a nucleus. As a result, the depth of Im$V_{K^-} \simeq(70 - 170)$~MeV in the KM1, P1, and M1 models 
and Im$V_{K^-} \simeq 270$~MeV in the KM2 and P2 models (not quoting uncertainties).  
The range of $V_{K^-}^{(2)}$ potential is considerably 
smaller than the range of the $V_{K^-}^{(1)}$ potential and thus in the surface region of a nucleus, $K^-$  single-nucleon absorption dominates in accordance with experimental findings~\cite{bubble1, bubble2, bubble3}.

The B2, B4, and M2 models yield the real part of the total $K^-$-nucleus potential extremely deep, 
$\sim~(200 - 300)$~MeV  in the nuclear interior, thanks to a strongly attractive Re$V_{K^-}^{(2)}$. On the contrary, the imaginary part of the $V_{K^-}$ potentials in these models 
is shallower than in the KM1 model.
\begin{figure}[t]
\begin{center}
\includegraphics[width=0.45\textwidth]{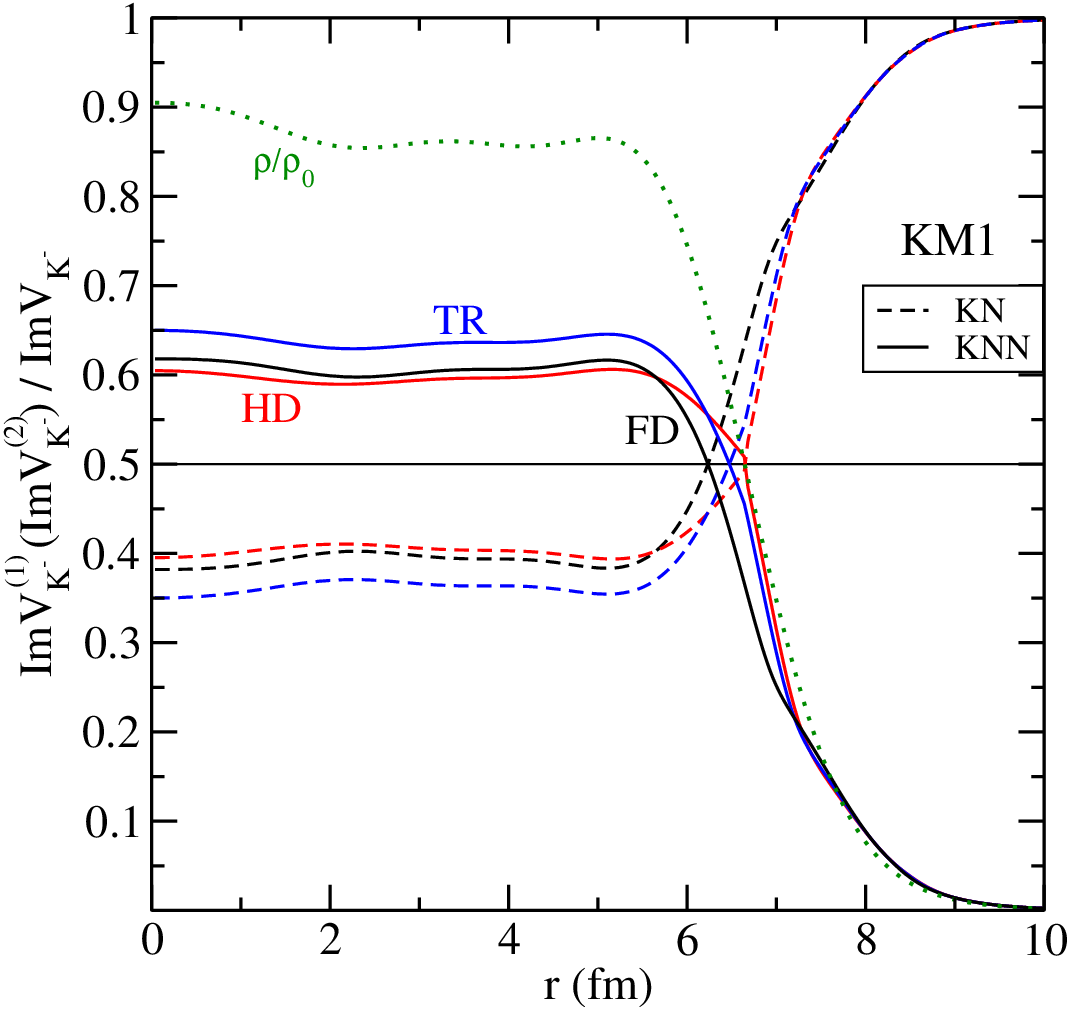}
\end{center}
\caption{The ratio of Im$V_{K^-}^{(1)}$ (dashed line) and Im$V_{K^-}^{(2)}$ (solid line) potentials to the total $K^-$ imaginary 
potential Im$V_{K^-}$ as a function of radius, calculated self-consistently for the $1s$ $K^-$ state in $^{208}$Pb using the KM1 model and 
different options for the $K^-$ multinucleon potential. The relative nuclear density $\rho / \rho_0$ (dotted line) is shown for comparison.}
\label{pomerknn}
\end{figure}

Next, we evaluated the fractions of the $K^-$ single- and multinucleon absorptions as a ratio of Im$V_{K^-}^{(1)}$ and Im$V_{K^-}^{(2)}$ with respect to the total imaginary 
$K^-$ potential Im$V_{K^-}$. These ratios are depicted in Fig.~\ref{pomerknn} as a function of radius, calculated self-consistently for the $1s$ $K^-$ state in $^{208}$Pb using the KM1 model and HD, TR, and FD options for $V_{K^-}^{(2)}$. For comparison, the relative density $\rho / \rho_0$ (thin dotted line) is shown here as well. Since the range and density dependence of $V_{K^-}^{(1)}$ and $V_{K^-}^{(2)}$ potentials is different (see Fig.~\ref{potknn}) the relative 
contribution of Im$V_{K^-}^{(1)}$ and Im$V_{K^-}^{(2)}$ to $K^-$ absorption is changing with the radius (density). In the surface region of a nucleus, the dominant 
process is the $K^-$ absorption on a single nucleon, while in the nuclear interior, the single-nucleon absorption is reduced due to the vicinity 
of the $\pi\Sigma$ threshold and multinucleon absorption prevails.
\begin{table}[h!]
\begin{center}
\caption{1s $K^-$ binding energies and widths (in MeV) in various nuclei calculated using the single nucleon 
$K^-N$ KM amplitudes (denoted KN); plus a phenomenological amplitude $B(\rho/\rho_0)^{\alpha}$, where $\alpha=1$ and 2, 
for 'half-density limit' (HD), $t\rho$ option (TR), and full density option (FD).}
\vspace*{8pt}

 \begin{tabular}{r c|c|c c c|c c c}
   \multicolumn{3}{c}{KM model}& \multicolumn{3}{|c|}{$\alpha = 1 $}& \multicolumn{3}{|c}{$\alpha = 2$} \\ \hline
   &  & KN & $\;$ HD & TR & FD & $\;$ HD & $\;\;$TR & FD \\ \hline 
 $^{6}$Li  & $B_{K^-}$ & 25 & $\; 11$ & not & not & $\; 23$ & $\; 19$ & not  \\[-5pt] 
      & $\Gamma_{K^-}$ & 45 & 116 & bound & bound & 122  & 160 & bound \\ \hline
 $^{12}$C  & $B_{K^-}$ & 45 & $\; 34$ & $\; 20$ & not & $\; 48$ & $\; 44$ & not  \\[-5pt] 
      & $\Gamma_{K^-}$ & 44 & 114 & 182 & bound & 125  & 191 & bound \\ \hline
 $^{16}$O  & $B_{K^-}$ & 45 & $\; 34$ & $\; 25$ & not & $\; 48$ & $\; 46$ & not  \\[-5pt] 
      & $\Gamma_{K^-}$ & 40 & 109 & 158 & bound & 121  & 167 & bound \\ \hline
 $^{40}$Ca  & $B_{K^-}$ & 59 & $\; 50$ & $\; 40$ & not & $\; 64$ & $\; 63$ & not  \\[-5pt]  
      & $\Gamma_{K^-}$ & 37 & 113 & 164 & bound & 126  & 175 & bound \\ \hline
 $^{90}$Zr  & $B_{K^-}$ & 69 & $\; 56$ & $\; 47$ & $\; 17$ & $\; 72$ & $\; 71$ & $\; 30$ \\[-5pt]   
      & $\Gamma_{K^-}$ & 36 & 107 & 156 & 312 & 120  & 167 & 499 \\ \hline
 $^{208}$Pb  & $B_{K^-}$ & 78 & $\; 64$ & $\; 56$ & $\; 33$ & $\; 80$ & $\; 80$ & 
$\; 53$  \\[-5pt] 
      & $\Gamma_{K^-}$ & 38 & 108 & 153 & 273 & 122  & 163 & 429 \\ \hline \hline 
   \multicolumn{3}{c}{P model}& \multicolumn{3}{|c|}{$\alpha = 1 $}& \multicolumn{3}{|c}{$\alpha = 2$} \\ \hline
   &  & KN & HD & TR & FD &  HD & TR & FD \\ \hline 
 $^{6}$Li  & $B_{K^-}$ & 38 & $\; 21$ & not & not & $\; 36 $ & $\; 28 $ & not  \\[-5pt]  
      & $\Gamma_{K^-}$ & 40 & 112 & bound & bound & 133  & 183 & bound \\ \hline
 $^{12}$C  & $B_{K^-}$ & 64 & 50 & $\; 35$ & not & $\; 64 $ & $\; 57 $ & not  \\[-5pt]  
      & $\Gamma_{K^-}$ & 28 & 96 & 165 & bound & 122  & 196 & bound \\ \hline
 $^{16}$O  & $B_{K^-}$ & 64 & 50 & $\; 39$ & not & $\; 63$ & $\; 59$ & not  \\[-5pt]  
      & $\Gamma_{K^-}$ & 25 & 94 & 142 & bound & 117  & 169 & bound \\ \hline
 $^{40}$Ca  & $B_{K^-}$ & 81 & 67 & $\; 56$ & not & $\; 82$ & $\; 79$ & not  \\[-5pt]  
      & $\Gamma_{K^-}$ & 14 & 95 & 145 & bound & 120  & 175 & bound \\ \hline
 $^{90}$Zr  & $B_{K^-}$ & 90 & 74 & $\; 62$ & $\; 19$ & $\; 87$ & $\; 85$ & not  \\[-5pt]  
      & $\Gamma_{K^-}$ & 12 & 88 & 136 & 340 & 114  & 164 & bound \\ \hline
 $^{208}$Pb  & $B_{K^-}$ & 99 & 82 & $\; 70$ & $\; 37$ & $\; 96$ & $\; 92$ & 
$\; 47^{*}$  \\[-5pt] 
      & $\Gamma_{K^-}$ & 14 & 92 & 137 & 302 & 117  & 163 & $412^{*}$ \\
\hline
\end{tabular}\label{bkgk2}
\end{center}
\vspace*{2pt}
\hspace*{25pt} $^*$ the solution of the Klein-Gordon equation for Im$V_{K^-}$ scaled by factor 0.8
\end{table} 
All three higher-density versions of $V_{K^-}^{(2)}$ yield the same fractions of single- and multinucleon absorption 
at the nuclear surface and differ slightly from each other inside the nucleus. 

The above discussed $K^- N$ amplitude models supplemented by $K^-$ multinucleon interactions described by 
the phenomenological potential $V_{K^-}^{(2)}$ were applied to calculations of $K^-$-nuclear bound states 
in various nuclei across the periodic table. We considered all three extrapolations HD, TR, and FD of 
$V_{K^-}^{(2)}$.\\     
In Table~\ref{bkgk2} we present $1s$ $K^-$ binding energies $B_{K^-}$ and widths $\Gamma_{K^-}$, calculated 
in the KM and P models, respectively. For comparison, we show also $K^-$ binding energies and widths calculated only   
with the underlying chirally-inspired $K^-$ single-nucleon potential. In these models which provide reasonable 
description of kaonic atom data and fractions of $K^-$ single- and multinucleon absorptions at rest, $K^-$ widths increase 
considerably after including $K^-$ multinucleon processes, while $K^-$ binding energies change only slightly 
(they decrease in KM1, P1, and P2\footnote{For the FD variant of the P2 model, we had to scale huge imaginary part Im$V_{K^-}$ by factor 0.8 in order to get fully converged self-consistent solution of the Klein-Gordon equation Eq.~\eqref{KG}. The calculation with the unscaled imaginary potential is not numerically stable due to extremely strong $K^-$ absorption --- the non-converged $\Gamma_{K^-} > 500$~MeV while the corresponding $B_{K^-} < 15$~MeV.} models and increase in KM2 model). 
For the FD multinucleon potentials $V_{K^-}^{(2)}$, the antikaon is unbound in the vast majority of nuclei. 
In $^{90}$Zr and $^{208}$Pb, we found $1s$ $K^-$ quasi-bound states, however, the $K^-$ binding energies of such 
states are small and widths are huge, one order of magnitude larger than the binding energies. 
For other variants of $V_{K^-}^{(2)}$ potential, 
HD and TR, $K^-$ widths are of order $\sim 100$~MeV but again, the binding energies are much smaller than the 
widths in most nuclei. The smallest $K^-$ widths are predicted in the P model for $\alpha=1$ and the HD option,  
nevertheless, they still exceed noticeably the binding energies. These results hold generally and remain valid even when the uncertainties in the multinucleon potential $V_{K^-}^{(2)}$ are taken into account.

\begin{table}[t!]
\begin{center}
\caption{1s $K^-$ binding energies and widths (in MeV) in $^{16}$O and $^{208}$Pb calculated using the single-nucleon 
$K^-N$ amplitudes M1, M2, B2, B4 plus a phenomenological amplitudes $B(\rho/\rho_0)^{\alpha}$ from Table~1.}
\vspace*{8pt}
 \begin{tabular}{r c| c c | c c | c c | c c }
   &  &  \multicolumn{2}{c|}{M1} & \multicolumn{2}{c|}{M2} & \multicolumn{2}{c|}{B2} & \multicolumn{2}{c}{B4}    \\ \hline
   & & $\;$ KN & FD & $\;$ KN & FD & KN & FD & KN & FD \\ \hline
 $^{16}$O  & $B_{K^-}$ & 25 &  $\; 48$ & 10 &135 & not &$\; 98$ & not & 170   \\ 
      & $\Gamma_{K^-}$ & 16 &117 & 22 &244 & bound &271 & bound & 190  \\ \hline
 $^{208}$Pb  & $B_{K^-}$ & 56 & $\; 80$ & 38 & 170 & not &146 & not & 200 \\ 
      & $\Gamma_{K^-}$ & 14 &121 & 32 & 214 & bound & 259 & bound & 174  \\ 
\end{tabular}\label{bkgk3}
\end{center}
\end{table} 

For completeness, we show in Table~\ref{bkgk3} binding energies and widths of the $K^-$ $1s$ states in 
$^{16}$O and $^{208}$Pb, calculated in M1, M2, B2, and B4 models and FD variant of  $V_{K^-}^{(2)}$. 
Unlike KM and P models, these models give $K^-$ quasi-bound states for the FD option also in $^{16}$O due to strongly attractive $K^-$ multinucleon interactions. 
However, the predicted $K^-$ binding energies are 
again much smaller than the widths (except the B4 model which yields comparable binding energies 
and widths). However, it is to be stressed 
that none of the models in Table~\ref{bkgk3} reproduces experimental values of the fractions of $K^-$ 
single- and multinucleon absorptions at rest. 

\begin{table}[h!]
\begin{center}
\caption{$K^-$ binding energies and widths (in MeV) in $^{208}$Pb+$K^-$ calculated using the single nucleon 
$K^-N$ KM amplitudes (denoted KN); plus a phenomenological amplitude $B(\rho/\rho_0)^{\alpha}$, where $\alpha=1$, 
for half density (HD) and full density (FD) options (see text for details).}
\vspace*{8pt}
 \begin{tabular}{c c|c c c c c c c}
   \multicolumn{2}{c|}{$^{208}$Pb+$K^-$} & $1s$ & $1p$ & $1d$ & $1f$ & $1g$ & $1h$ & $1i$ \\ \hline 
 KN  & $B_{K^-}$ & 78  & 70 & 61 & 52 & 42 & 31 & 20  \\ 
      & $\Gamma_{K^-}$  & 38  & 38 & 40 & 42 & 45 & 46 & 47 \\ \hline
 HD  & $B_{K^-}$ & 64  & 58  & 51 & 42 & 33 & 22 & 8 \\ 
      & $\Gamma_{K^-}$  & 108 & 110 & 112 & 115 & 120 & 127 & 143 \\ \hline
 FD & $B_{K^-}$  & 33  & 24 & 9  & not & not & not & not  \\ 
      & $\Gamma_{K^-}$ & 273 & 285  & 306 & bound & bound & bound & bound \\ \hline
\end{tabular} \label{bkgk1}
\end{center}
\end{table} 

Table~\ref{bkgk1} presents the binding energies and widths of $K^-$ quasi-bound states in $^{208}$Pb, calculated in the KM1 model with FD and HD options of the multinucleon potential. The binding energies and widths of $K^-$ states calculated with the underlying $K^-N$ single-nucleon potentials (KN) are presented here for comparison. The $K^-N \rightarrow \pi Y$ conversion widths are gradually increasing in excited states as $\delta \sqrt{s}$ is moving away from the $\pi \Sigma$ threshold. However, the increase in the KM model is not as pronounced as in the P model \cite{gmNPA} where the difference between the $K^-$ widths due to $K^-$ single-nucleon absorption in the $1s$ and $1i$ states is $\simeq 35$~MeV (compare also $\Gamma_{K^-}$ of excited states in $^{40}$Ca for various $K^-N$ amplitude models in Fig.~\ref{BkGkCa}). For the HD option of multinucleon potential, the $K^-$ binding energies are smaller and widths are more than twice larger than in the KN case. In the FD version of $V_{K^-}^{(2)}$, the number of excited $K^-$ quasi-bound states is considerably reduced because of strong $K^-$ absorption.

\section{Conclusions}
\label{sec-3}
We performed calculations of $K^-$ nuclear quasi-bound states using $K^-$-nucleus optical potentials derived
self-consistently from $K^- N$ amplitudes, obtained within several recent chirally-motivated
meson-baryon coupled-channel models.
Following analyses of Friedman and Gal \cite{fgNPA, fgNPA16} these models need to be supplemented by a
phenomenological term representing $K^-$ multinucleon interactions in order to fit kaonic atom data.
Though only the P and KM models are able to reproduce at the same time
the experimentally determined fractions of $K^-$ single-nucleon absorption
at rest~\cite{fgNPA16}, we considered also the other $K^- N$ amplitude models in order to explore model 
dependence of our calculations. The main aim of our work was to assess the effect of the $K^-$ multinucleon 
processes on binding energies and widths of kaonic nuclear states.

 First, we constructed the chirally-motivated $K^-$ single-nucleon part of the optical potential using
6 different sets of $K^-N$ amplitudes. In order to account for Pauli
correlations in the nuclear medium, we applied the multiple-scattering WRW procedure~\cite{wrw}. We verified
that hadron self-energies, considered in previous calculations of in-medium $K^-N$
amplitudes~\cite{cfggmPLB, cfggmPRC11}, affect the $K^-$ single-nucleon
potential only slightly in the energy region relevant to our current calculations. An important aspect of
chirally-motivated $K^-N$ amplitudes is their energy dependence which has to be treated self-consistently, taking
into account the non-negligible contribution from $K^-$ and $N$ momenta.
Each out of the considered models gives different depths of Re$V_{K^-}$ in a nucleus and thus probes different
energy regions below the $K^- N$ threshold. The resulting $K^-$ binding energies $B_{K^-}$  are then strongly model
dependent. The widths of the $1s$ $K^-$-nuclear states come out quite small. The smallest
widths $\Gamma_{K^-}$ are predicted by the Murcia model M1, whereas the KM model predicts the $K^-$ widths
three times as large.

Next, we added to each $K^-$ single-nucleon potential $V_{K^-}^{(1)}$ a corresponding phenomenological
multinucleon potential $V_{K^-}^{(2)}$, parameters of which were recently fitted to kaonic
atom data \cite{fgNPA16}.
Since the kaonic-atom data probe the $K^-$ optical potential reliably up to at most $\sim 50\%$ of $\rho_0$,
we considered three different scenarios for extrapolating $V_{K^-}^{(2)}$ to higher densities,
$\rho \geq 0.5\rho _0$. Though the applied models differ widely in the subthreshold region, our
calculations lead to some quite general conclusions, valid for each of the $K^-$-nucleus interaction
models. We found that the $K^-$ multinucleon absorption gives rise to substantial
increase of the widths of $K^-$-nuclear states. The $K^-$ widths exceed considerably the $K^-$ binding 
energies in the vast majority of nuclei.
In the KM and P models, the only models accepted by the analysis of Friedman and Gal \cite{fgNPA16},
the FD variant of $V_{K^-}^{(2)}$ even does not yield
any $K^-$-nuclear bound state in most of the nuclei under consideration.
We verified that these conclusions remain valid even after taking into account the uncertainties in the
multinucleon potential $V_{K^-}^{(2)}$.

After exploring various chirally-inspired coupled-channel models of meson-baryon interactions together
with a phenomenological $K^-$ multinucleon part fitted to reproduce the experimental data we feel free to
conclude that the widths of $K^-$-nuclear quasi-bound states in nuclei with $A \geq 6$
are considerably larger than their binding energies. Therefore, observation of such states
in experiment seems highly unlikely. We believe that our results will stimulate theoretical studies of the role of $K^-$ multinucleon processes in lighter $K^-$-nuclear systems in which few-body techniques are applicable.

\section*{Acknowledgements}
We wish to thank E. Friedman and A. Gal for valuable discussions, and A. Ciepl\'{y} 
and M. Mai for providing us with the free $K^-N$ scattering amplitudes.
This work was supported by the GACR Grant No. P203/15/04301s.\\  
J. H. acknowledges support from the CTU-SGS Grant No. SGS16/243/OHK4/3T/14.  
Both J. H. and J. M. acknowledge the hospitality extended to them at the Racah Institute of Physics, 
The Hebrew University of Jerusalem, during their collaboration visit in November 2016. 
J. M. acknowledges financial support of his visit provided by the Racah Institute of Physics. 
J. H. acknowledges financial support of the Czech Academy of Sciences which enabled her stay at the Hebrew University.

\end{document}